\documentclass[modern]{aastex63}
\pdfoutput=1

\begin{document}
\newcommand{\vdag}{(v)^\dagger}
\newcommand\aastex{AAS\TeX}
\newcommand\latex{La\TeX}
%-------------------------------
\newcommand\degree{degree}
\newcommand\kepler{\textit{Kepler} }
\newcommand{\ik}{{\it Kepler}}
\newcommand{\KIC}{\rm Kepler-453}
\newcommand{\etal}{{\it et~al.~}}
\newcommand{\ie}{{\it i.e.~}}
\newcommand{\eg}{{\it e.g.~}}
\newcommand{\vs}{{\it vs.~}}
\newcommand{\kms}{\mbox{$\rm km \ s^{-1}$}}
\newcommand{\Msun}{\mbox{M$_{\sun}$}}
\newcommand{\Mjup}{\mbox{M$_{Jup}$}}
\newcommand{\Mearth}{\mbox{M$_{\earth}$}}
\newcommand{\Rsun}{\mbox{R$_{\sun}$}}
\newcommand{\Rjup}{\mbox{R$_{Jup}$}}
\newcommand{\Rearth}{\mbox{R$_{\earth}$}}
\newcommand{\Lsun}{\mbox{L$_{\sun}$}}
\newcommand{\ltsimeq}{\raisebox{-0.6ex}{$\,\stackrel
        {\raisebox{-.2ex}{$\textstyle <$}}{\sim}\,$}}
\newcommand{\gtsimeq}{\raisebox{-0.6ex}{$\,\stackrel
        {\raisebox{-.2ex}{$\textstyle >$}}{\sim}\,$}}
\def\lesssim{\mathrel{\hbox{\rlap{\hbox{\lower4pt\hbox{$\sim$}}}\hbox{$<$}}}}
\def\gtrsim{\mathrel{\hbox{\rlap{\hbox{\lower4pt\hbox{$\sim$}}}\hbox{$>$}}}}
\def\ggg{\mathrel{\hbox{\rlap{\hbox{\lower4pt\hbox{$\sim$}}}\hbox{$>$}}}}
\def\persec{s$^{-1}$}
%-------------------------------

% ---------------------------------------------------------------------
\shorttitle{Kepler-1661 \MakeLowercase{b}:
A \textit{Kepler} Transiting CBP}
\shortauthors{Socia et al.}

% ---------------------------------------------------------------------

\title{Kepler-1661 \lowercase{b}: A Neptune-sized \textit{Kepler} Transiting Circumbinary Planet 
around a Grazing Eclipsing Binary\footnote{Based on observations obtained with the Hobby-Eberly Telescope, which is a joint project of the University of Texas at Austin, the Pennsylvania State University, Ludwig-Maximilians-Universit\"at M\"unchen, and Georg-August-Universit\"at G\"ottingen.}}

% --------------------------------------------

\author[0000-0002-7434-0863]{Quentin J Socia}
\affiliation{Department of Astronomy, San Diego State University, 5500 Campanile Drive, San Diego, CA 
92182-1221, USA}

\author[0000-0003-2381-5301]{William F Welsh}
\affiliation{Department of Astronomy, San Diego State University, 5500 Campanile Drive, San Diego, CA 
92182-1221, USA}

\author[0000-0001-9647-2886]{Jerome A Orosz}
\affiliation{Department of Astronomy, San Diego State University, 5500 Campanile Drive, San Diego, CA 
92182-1221, USA}

\author[0000-0001-9662-3496]{William D Cochran}
\affil{McDonald Observatory, The University of Texas as Austin, Austin, TX 
78712-0259, USA}

\author{Michael Endl}
\affil{McDonald Observatory, The University of Texas as Austin, Austin, TX 
78712-0259, USA}

\author[ 0000-0002-9644-8330]{Billy Quarles}
\affil{Center for Relativistic Astrophysics, School of Physics,
Georgia Institute of Technology,
Atlanta, GA 30332, USA}

\author{Donald R Short}
\affiliation{Department of Astronomy, San Diego State University, 5500 Campanile Drive, San Diego, 
CA 92182-1221, USA}

\author[0000-0002-5286-0251]{Guillermo Torres}
\affiliation{Center for Astrophysics $\vert$ Harvard \& Smithsonian, 60 Garden Street, 
Cambridge, MA 02138, USA}

\author[0000-0002-6742-4911]{Gur Windmiller} 
\affiliation{Department of Astronomy, San Diego State University, 5500 Campanile Drive, San Diego, 
CA 92182-1221, USA}

\author{Mitchell Yenawine} 
\affiliation{Department of Astronomy, San Diego State University, 5500 Campanile Drive, San Diego, 
CA 92182-1221, USA}

% ---------------------------------------------------------------------

\begin{abstract}
We report the discovery of a Neptune-size ($R_p=3.87\pm0.06 R_\oplus$) transiting circumbinary planet, 
{\mbox{Kepler-1661~b}}, found in the \kepler photometry.
The planet has a period of $\sim$175 days and its orbit precesses
with a period of only 35 years. The precession causes the alignment 
of the orbital planes to vary, and the planet is in a transiting 
configuration only $\sim$7\% of the time as seen from Earth.
As with several other \kepler \ circumbinary planets, Kepler-1661~b orbits
close to the stability radius, and is near the (hot) edge of habitable zone.
The planet orbits a single-lined, grazing eclipsing binary, containing a 
0.84 \Msun \ and 0.26 \Msun \ pair of stars in a mildly eccentric 
($e$=0.11), 28.2-day orbit. The system is fairly young, with an 
estimated age of $\sim$ 1-3 Gyrs, and exhibits significant starspot 
modulations. The grazing-eclipse configuration means the system is very 
sensitive to changes in the binary inclination, which manifests itself 
as a change in the eclipse depth. The starspots contaminate the eclipse 
photometry, but not in the usual way of inducing spurious eclipse timing 
variations. Rather, the starspots alter the normalization of the light 
curve, and hence the eclipse depths. This can lead to spurious eclipse 
depth variations, which are then incorrectly ascribed to binary orbital 
precession.

\end{abstract}


% ---------------------------------------------------------------------

\section{Introduction} \label{sec:intro}

If an exoplanet orbits two stars instead of one, it complicates the detection, 
characterization, and long-term behavior of the system. Yet, it also provides 
the opportunity to measure the stellar and planetary properties with exquisite 
precision. If the host stars are eclipsing and are bright enough for both 
their radial velocities to be measured, then we can directly determine the 
stellar masses, radii, temperatures, and age using the traditional binary star 
analysis techniques. If in addition the exoplanet transits the stars, then 
much more information about the binary system is available:
the transit times tell us 
about the relative locations of the bodies, and the transit durations
tell us about the relative velocities. 
The transit depths provide further information about the precession of the 
orbits and place much tighter constraints on the limb darkening and system parameters. 
It is no surprise that the circumbinary 
planet (CBP) systems have among the most accurate and precisely known stellar 
and planet parameters, e.g., the masses and radii of the stars in Kepler-34
are known to better than 0.3\%, and the planet's radius to 1.7\% 
\citep{2012Natur.481..475W};
in Kepler-16, the planet's radius is known to an astonishing 0.35\% 
\citep{Doyle_2011}.
Even when the secondary star's radial velocity is not measurable, the full set 
of system parameter can still be determined. The transits provide the 
information necessary to determine the binary mass ratio, something impossible 
in a classical single-lined eclipsing binary system.\footnote{We use the term ``eclipse" to refer to mutual star-star crossings, and the term ``transit" for a planet crossing in front of a star. No occultations are seen in Kepler-1661.} 
See the review by \citet{Welsh2018} for more on this topic.

However, such richness comes with a cost. The orbital motion of the planet is 
decidedly non-Keplerian, so the equations of motion need to be numerically 
integrated, including corrections for general relativity and apsidal motion. 
More importantly, the standard technique for exoplanet mass determination -- 
measurement of the Doppler reflex motion of the host star -- has not yet 
worked for a CBP. 
The meters-per-second radial velocity induced by the planet is completely 
dwarfed by
the much shorter timescale and larger amplitude 
velocity variation caused by the companion star. 
Fortunately, there is a way to determine the planet's mass: the planet induces 
variations in the eclipse times. 
The larger the planet's mass, the larger the eclipse timing variations (ETVs). 
The orbital period, eccentricity, and argument of periastron can be measured, 
and the planet need not be transiting to be detected:
the orbital inclination can be constrained given high-enough quality data. 
See \citet{Borkovits2011} and \citet{Borkovits2015} for a full discussion of the ETV method. 
Note that eclipses are usually much deeper than transits, and therefore the 
eclipse times can be measured with very high precision, even with ground-based 
photometry. An eclipse timing uncertainty of 10 seconds in a 30-day binary 
amounts to a precision of 4 ppm. So even though the planet barely perturbs the 
binary, at this precision its presence can be felt. Unlike radial velocities, 
eclipse timing variations due to precession grow with time, so a long temporal 
baseline can more than compensate for less-than-\textit{Kepler}-quality 
observations.

KIC 6504534 was discovered and cataloged as a $\sim$28.2 day 
eclipsing binary system in the second revision of the \kepler 
Eclipsing Binary Catalog\footnote{http://keplerebs.villanova.edu/} 
(\citealt{Prsa2011}; \citealt{Slawson2011}).
At the time the binary was discovered, the planet was not transiting. 
The first transit (near BJD 2455804.8, 2011 Aug 11 UT) did not occur 
until Quarter 10. Visual inspection of the light curve revealed a 
second transit near BJD 2455975.1 (2012 Feb 17) in the Quarter 12 data, 
allowing a rough estimate of 180 days to be made for the candidate's orbital 
period. 
The target was requested to be observed in Short Cadence mode 
(approximately 2 minute sampling instead of 30 minutes) and the planet host
candidate was given the designation KOI-3152. A third transit was observed 
in Quarter 14 (BJD 2456145.5, 2012 Aug 05), but sadly a fourth transit 
event that occurred in Quarter 16 fell in a gap and was not observed.

KOI-3152, now known as Kepler-1661, is a single-lined eclipsing binary, so the mass ratio cannot be 
determined from the radial velocities alone. While in principle 
transits can provide enough information to determine the mass ratio (e.g.\ 
this was done for Kepler-16 \citep{Doyle_2011}), the three transits in 
Kepler-1661 all occur near the same binary orbital phase (phase 0.40, 0.45, 0.5) 
and thus provide only limited information on the primary star's orbit. This 
fundamentally limits our ability to precisely measure the system parameters. 
Nevertheless, there is enough information to fully characterize the binary, and 
photodynamical modeling clearly establish the candidate as a CBP. 
In Section \ref{sec:2} we present the observations, and in Section \ref{sec:pdm} review the photodynamical 
modeling at length, with particular emphasis on treatment of the effects of 
starspots in the light curve. In Section \ref{sec:discuss} we discuss the results and properties
of the binary star and new circumbinary planet.

% ----------------------------------------------------------------------------

\section{Observations} \label{sec:2}

\subsection{Kepler Data}\label{sec:kep}

\begin{figure}[ht]
  \centering
    \includegraphics[width=1\textwidth]{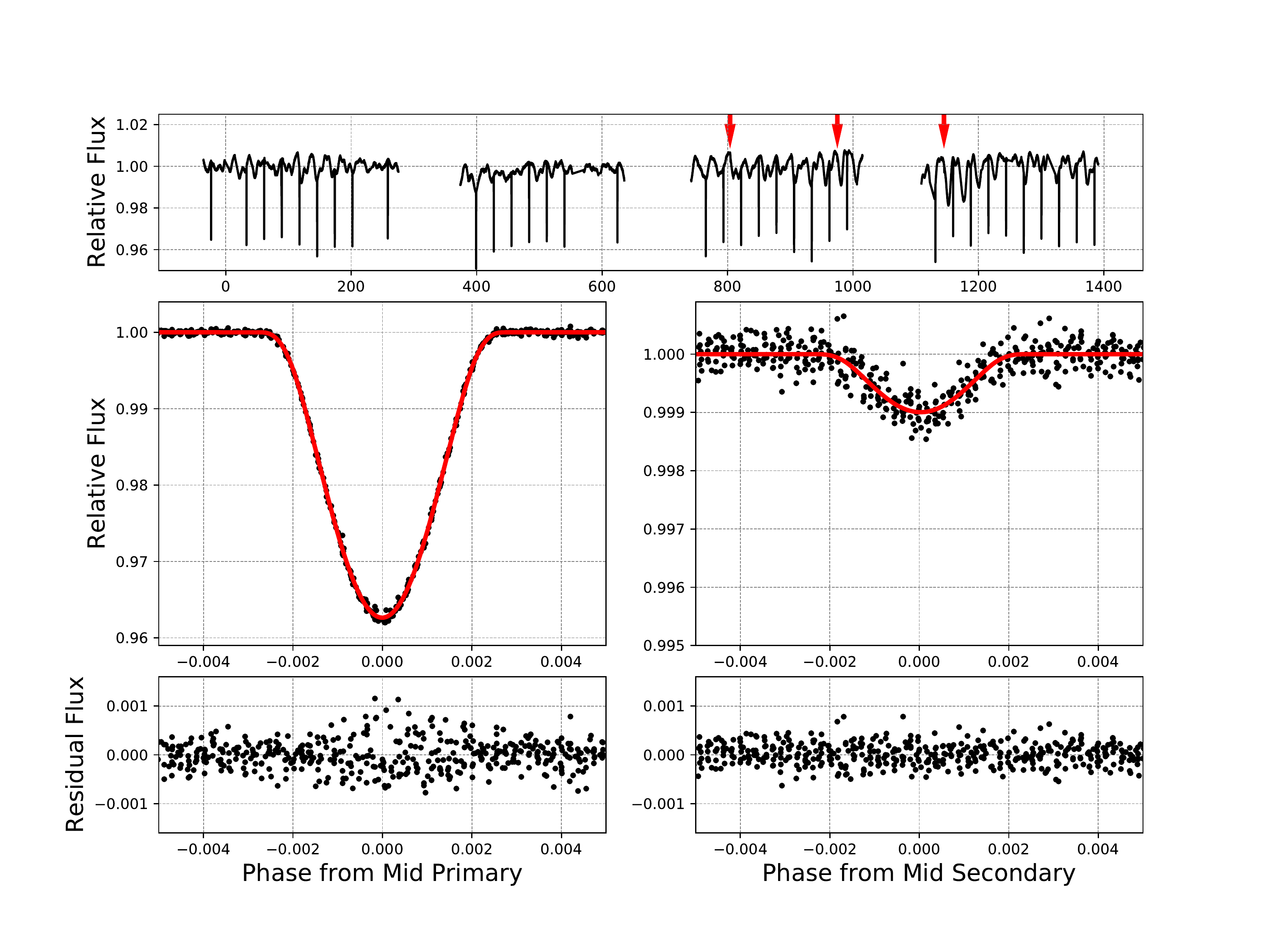}
  \caption{
The normalized {\it Kepler} light curve of Kepler-1661 is shown in the upper panel with the abscissa representing BJD-2455000.
Each Quarter was detrended with a cubic polynomial.
The red arrows indicate where the three transits of the planet occur. 
The lower panels show the orbital phase-folded primary and secondary eclipses, 
along with an initial model fit and residuals. The V-shaped primary eclipse
immediately tells us that the eclipse is grazing.
The residuals show a larger scatter during eclipse than outside eclipse, 
most likely due to changes in the normalized eclipse depth caused by starspots.
  \label{fig:bin}}
\end{figure}

All available data on Kepler-1661 were retrieved from MAST in 
early 2019, corresponding to {\it Kepler} Data Release DR25. We use the SAP 
(Simple Aperture Photometry) calibration, not the PDC-MAP calibration, 
because we find it preserves intrinsic stellar variability with higher 
fidelity. Kepler-1661 fell on one of the failed CCD modules (module 3) 
and thus no observations are available for Quarters 5, 9, 13 and 17. This 
results in three $\sim$ 90-day gaps in the light curve which otherwise is 
superb, as is typical of {\it Kepler} photometry.
Quarters 14, 15, and 16 have Short Cadence data available, and these were 
used in the preliminary investigation and for building a template eclipse profile 
for the eclipse timing measurements. However, because of the relative faintness 
of the target, these were not used in the final photodynamical modeling.
The upper panel of Figure \ref{fig:bin} shows the entire {\it Kepler} light curve. 
For this figure, each Quarter was detrended and normalized using a third 
order polynomial. The primary eclipses are readily seen, as are ripples
due to starspots.
A total of 36 primary eclipses are present, although 
two are missing ingress data and are mostly unusable.
The primary eclipses have a fractional depth of $\sim$ 0.038; in contrast, the 
36 secondary eclipses are not even visible on this scale, having a fractional depth of only
$\sim$0.001. 
Also shown in Figure \ref{fig:bin} are the phase-folded eclipse profiles.
Before folding on the orbital period, each eclipse was detrended and
normalized by masking out the eclipse and fitting a 3rd order polynomial 
to a narrow window surrounding each eclipse, and the data were 
then divided by the polynomial fit.
The shallow depths and V-shaped eclipses are indicative of grazing
eclipses. The residuals of an initial model fit to the eclipses are
flat for the secondary, but show an increased scatter during the
primary eclipse. This indicates a change in eclipse depth, potentially
due to starpots.

Also present in the light curve are three transits, and notably, 
the transit depths increase from $\sim$ 0.0018 to 0.0023 
as shown in Figure \ref{fig:trans}.
This change in transit depth implies a change in the inclination
(or impact parameter) of the planet's orbit, a consequence of
rapid precession. The transit widths also increase, 
consistent with a decrease in the impact parameter, though
a change in transit duration can also be caused by a change in 
the relative velocity of the planet and star at different orbital
phases of the star.
During the third transit ingress there is a datum missing, and 
the {\it Kepler} pipeline Data Quality flag indicates a cosmic ray hit 
occurred on the CCD column at this time. The ingress looks by eye to be
fine, but to be cautious, we boosted the uncertainty on the data 
points after this cosmic ray event by a factor of 10.

The {\it Kepler} Input Catalog (KIC) provides the following estimates 
for the stellar parameters: 
Kepmag = 14.216, 
$T_{eff}$ = 4748 K,
$\log{g}$ = 4.46,
metallicity = -0.10,
and a contamination = 0.00 for all four {\it Kepler} Seasons.
In general, the KIC estimates should be used with considerable caution
for binary stars, but in this case the primary star dominates the light 
from the system (see Section \ref{sec:binary}) so these estimates should not be heavily 
biased.

% --------------------------------------------
\subsection{Mt.\ Laguna Observatory Photometry}

A primary eclipse on 2019 Jun 03 (UT) was observed in the Johnson-Cousins 
\textit{R}-band with the Mount Laguna Observatory (MLO) 1-meter telescope.
Exposures of 120 seconds were used and the CCD pixels were binned 2 x 2 
during readout, giving an effective pixel size of 0.8 arcseconds, more than 
adequate for the relatively poor 3-arcsecond average seeing that night.
Standard data calibration was performed using AstroImageJ (AIJ; 
\citealt{Collins2017}), and differential photometry carried out
using six comparison stars within 3 arcminutes of the target. 
AIJ utilizes the UTC2BJD calculator \citep{2010PASP..122..935E} 
to convert UT to BJD times.

\begin{figure}[ht]
  \centering
    \includegraphics[width=1\textwidth]{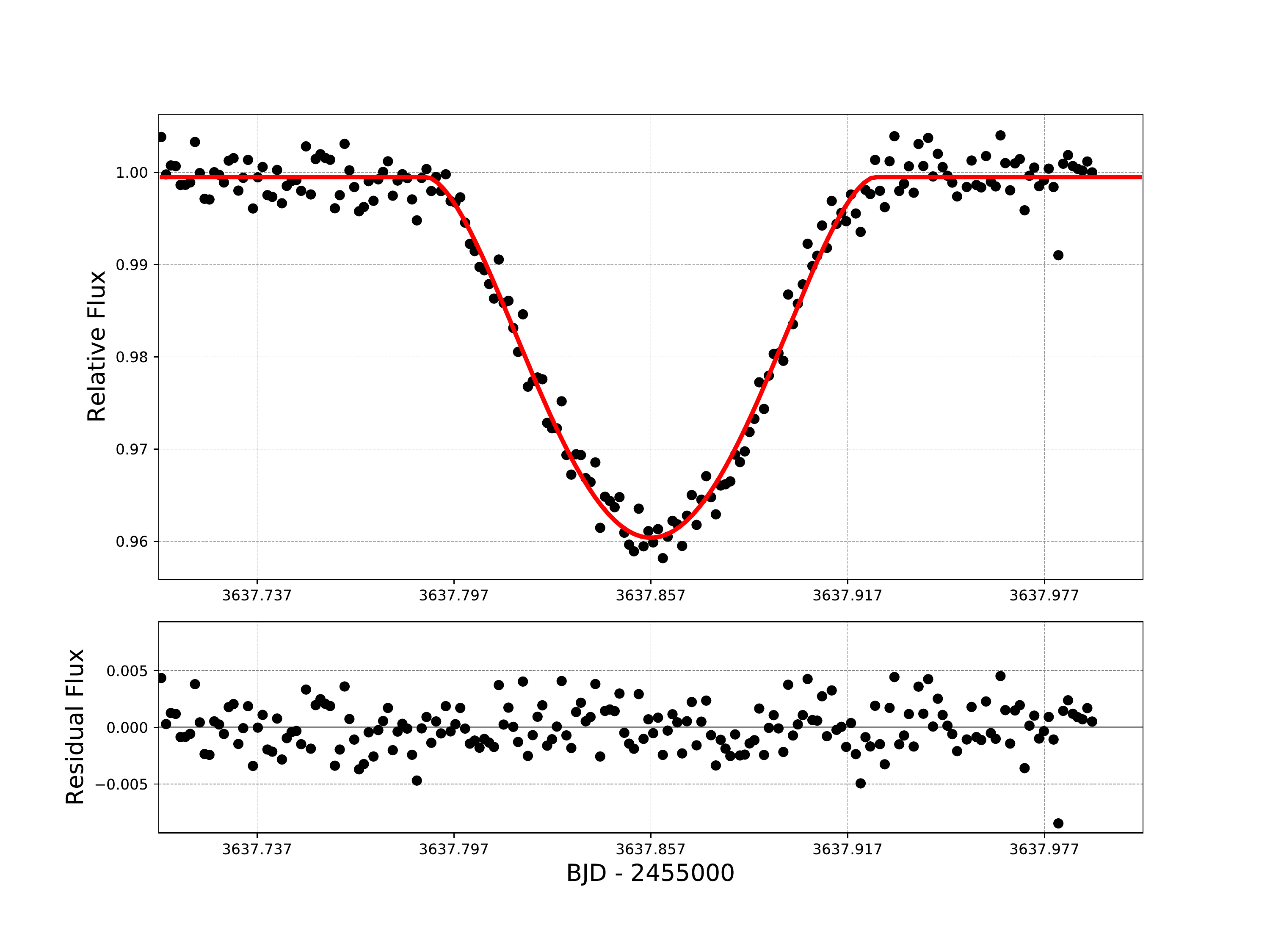}
  \caption{
Normalized and detrended MLO R-band primary eclipse with the best-fit 
photodynamical model shown in red. The eclipse depth is slightly deeper than the \textit{Kepler} eclipses because of the wavelength-dependence of the limb darkening combined with the high impact parameter of the grazing eclipse.
  \label{fig:mlo}}
\end{figure}

The time interval from the first to the last primary eclipse in the 
\kepler light curve was 1407 days. The eclipse provided by MLO more 
than doubles the temporal baseline, extending it to 3660 days (10 years).
Figure \ref{fig:mlo} shows the MLO light curve. 
A mid-eclipse time of 2458637.8570 $\pm$ 0.0003 BJD was measured, 
which is within 1$\sigma$ of the extrapolated linear ephemeris determined 
from the \kepler primary eclipses. It should be noted that due to the different bandpass, and hence limb darkening, the R-band eclipse shape and depth are different from that seen by \textit{Kepler}.

% --------------------------------------------
\subsection{Spectroscopy and Radial Velocities}

We obtained high-resolution spectra with three instruments: 
the HRS spectrograph on the Hobby-Eberly Telescope (HET; \citealt{tull1998}), 
the Tull Coude spectrograph on the 2.7-m Smith telescope at McDonald 
Observatory \citep{tull1995}, and the echelle spectrograph on the 4-m Mayall 
telescope at Kitt Peak National Observatory (KPNO).  The radial velocity standard 
star HD~182488 was observed with each 
spectrograph to assist in calibration of the velocity zero-point. The HET and 2.7m Smith telescope observations used Coude spectrographs and made targeted observations of Kepler-1661. The Mayall telescope observations used a Cassegrain spectrograph and the observations were made as part of a survey of Kepler eclipsing binaries. The former observations have higher accuracy than the latter.

\begin{figure}[ht]
  \centering
    \includegraphics[width=1\textwidth]{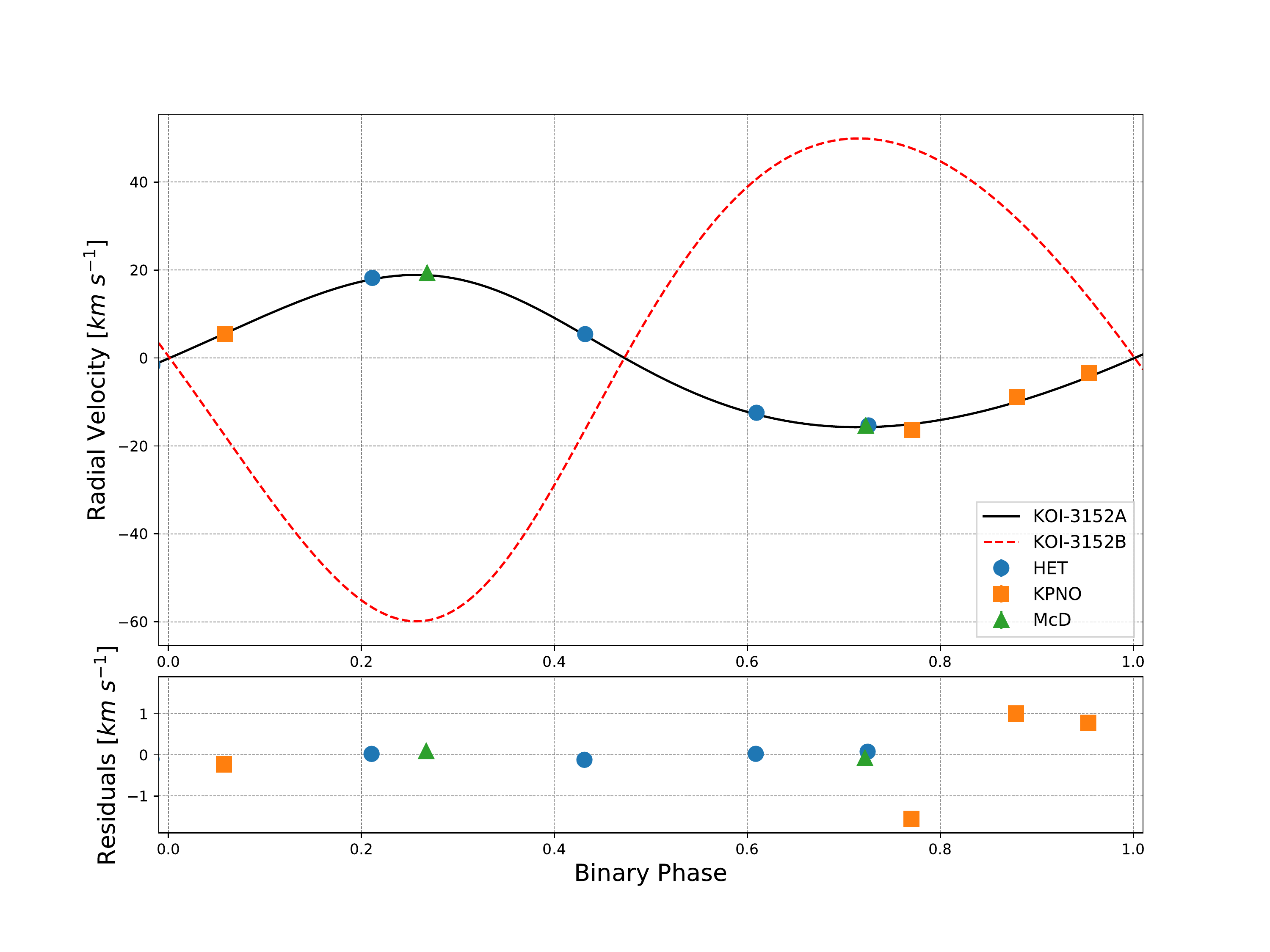}
  \caption{Radial velocities of the primary star with the best model fit folded on the binary period. 
The uncertainties in the velocities are smaller than the symbols.
The secondary star is not detected in the spectra, but its expected 
radial velocity curve is shown as the red dashed curve.
  \label{fig:rvs}}
\end{figure}

A total of 11 radial velocities 
were obtained over the course of 2012 and 2013, and are listed in Table 
\ref{tab:rvs}. Only the primary star's spectrum was detectable in the spectra, 
consistent with the very shallow secondary eclipses. The radial velocity curve 
is shown in Figure \ref{fig:rvs} and is well-matched by our photodynamical 
model fit (discussed in Section \ref{sec:revised}) with a radial velocity semi-amplitude ($K$) of 17.3 \kms \ and 
eccentricity 0.11. Nearly identical values are obtained when just the radial 
velocities alone are fit with a simple binary star model.

Using the Kea code \citep{2016PASP..128i4502E} 
on the 2.7-meter observations (these have the highest spectral resolution,
R$\sim$ 60,000) yields a mean effective temperature of 
% 5100 $\pm$ 75 K,
5140 $\pm$ 50 K, 
metallicity [M/H] = -0.12 $\pm$ 0.10 dex, 
log(g) = 4.66 $\pm$ 0.10 cgs, 
and $V_{\rm{rot}} \sin{i}$ = 2.5 $\pm$ 0.5 \kms.

For an independent estimate of the temperature, we used published photometry 
to compute 5 color indices, dereddened using 4 different dust maps
(giving E(B-V) = 0.056 for an assumed distance of 400 pc).
The \citet{2010A&A...512A..54C} color/temperature transformations
yield an average $T_{eff}$ of 5070 $\pm$ 110 K, assuming solar metallicity and 
that the secondary star does not contribute a significant amount of light.
This is in excellent agreement with the spectroscopic value. As a final value,
we adopt an effective temperature for the primary star of 5100 $\pm$ 100 K.

% --------------------------------------------
\subsection{Gaia Parallax}

Gaia Mission data for Kepler-1661 (Gaia DR2 2104078025612319360) were retrieved 
from the Gaia Data Release 2 
(\citealt{2016A&A...595A...1G}; \citealt{2018A&A...616A...1G}).
Gaia measured a parallax of 2.407 $\pm$ 0.017 mas,
giving a distance of 415 $\pm$ 3 parsecs.
With this distance, plus the precise absolute Gaia photometry and the 
effective temperature, we can estimate the radius of the primary star, 
assuming the secondary star contributes a negligible amount of light. 
This assumption is consistent with the lack of detection of the secondary 
in the spectroscopy and the shallow eclipse depth in the photometry. It is 
later shown to be completely justified by the photodynamical model.

We used the conversion given by the Gaia Collaboration to estimate 
the bolometric correction to the \textit{G}-band magnitude of 14.19 
(with an estimated error of 0.02 mag) and an extinction of 
$E(B-V) = 0.056 \pm 0.02$.  We derive a radius of
0.743 $\pm$ 0.042 \Rsun\ for the primary star. The uncertainty in the
temperature is the dominant source of the uncertainty, but the uncertainty
in the reddening is a major contributor.

% --------------------------------------------
\subsection{Photometric Contamination}
With 4-arcsecond pixels and an aperture several pixels across, the potential 
for unwanted light to be included in the \kepler photometry is significant. 
Since this extra light has the effect of reducing the eclipse and transit depths, 
and thus the stellar and planetary radii (and other correlated parameters), 
it is important to constrain the excess light contamination as much as possible. 
The MAST website gives a contamination level of zero for all Quarters, except 
Quarter 1 which has a contamination of 0.0004 listed. 

To verify this, we queried the Gaia catalog for any objects within 40 arcseconds 
of Kepler-1661 and found that of the 15 objects returned, only one was sufficiently 
close and bright enough to possibly contaminate the \kepler photometry:
Gaia DR2 2104078025609950464, which is KIC~6504533.
Located $\sim$9 arcseconds away, it is 6.55 magnitudes dimmer in the 
Gaia \textit{G} bandpass \citep{Riello2018}  and if this star were 
{\it entirely} within the \kepler aperture it would account for only 
$\sim$0.003 of the observed flux. But in fact this star does not lie within the 
\kepler aperture for any Quarter.

Aligning and stacking together our MLO R-band data 
into one image ($\sim$7-hour exposure),
we confirm the nearby star's location and brightness.
This star is well outside of the aperture used to generate the MLO light curve,
and we found no other sources of light near Kepler-1661.
Finally, we examined the \kepler Target Pixel Files and found no indication of 
additional background light, nor any image centroid movement during eclipse.
Thus the very low level of contamination listed at MAST seems correct and
we treat the contamination as negligible for all Quarters.

% ---------------------------------------------------------------------------------
\section{Photodynamical Modeling} \label{sec:pdm}

% --------------------------------------------
\subsection{The ELC Photodynamical Model}

We performed a simultaneous fit of all the eclipses and transits along with 
the radial velocity measurements using the eclipsing light curve code ``ELC'' 
(\citealt{2000A&A...364..265O}; \citealt{Wittenmyer2005}; \citealt{Orosz2019}). 
The code integrates the equations of motion using Newtonian 
gravity with general relativistic corrections (\citealt{2002ApJ...573..829M}; \citealt{2009ApJ...698.1778R}; \citealt{2001icbs.book.....H}). A ``tidal'' term is also included 
to account for the non-spherical potentials, which leads to classical apsidal 
motion, but this effect is negligible in comparison to the dynamical and 
general relativistic precession (which themselves are small). A 12$^{th}$ 
order Gaussian Runge-Kutta symplectic integrator is used, based on the code 
of \citet{Hairer}. We employed both a nested sampling algorithm
\citep{Skilling2004}
and
a Differential Evolution Monte Carlo Markov Chain (DE-MCMC) technique to 
sample the posterior distribution of the parameters 
\citep{tbv2006} and estimate their uncertainties.

The model stellar eclipses and planet transits are computed using the method 
outlined in \citet{2018AJ....156..297S}, replacing the \citet{2002ApJ...580L.171M} and 
Gim\'{e}nez methods \citep{2006A&A...450.1231G} formerly used in ELC. A quadratic limb 
darkening law is used, following the prescription of \citet{2013MNRAS.435.2152K} to more 
efficiently sample the correlated limb darkening coefficients. A total of 25
parameters are used in the model: the five standard Keplerian orbital 
parameters for each orbit ($P,T_{c},i,e,\omega$), the masses and radii of the 
three bodies, the stellar temperatures, two quadratic limb darkening 
coefficients for each star in the {\it Kepler} and R-bandpasses, and the 
longitudinal nodal angle $\Omega_p$ of the planet's orbit ($\Omega_b$ of the 
binary is set fixed to zero). 
In the actual fitting 
procedure, ratios and other combinations of parameters are often better 
constrained by the data or better sampled by the DE-MCMC process and therefore 
these equivalent parameters are used, e.g.\ orbital velocity of the primary 
$K_{1}$, mass ratios, radius ratios, temperature ratio, 
$\sqrt{e} \cos{\omega}$, and $\sqrt{e} \sin{\omega}$. The temperature of the 
primary star is also a free parameter though the light curves and radial 
velocity observations cannot constrain it. The spectroscopically 
determined value and its uncertainty are included as a datum that the model 
needs to match (i.e.\ it is included the $\chi^{2}$ statistic). The same 
procedure is used to steer the solutions towards the primary star radius
determined with the Gaia parallax. The radius is free to be any value but 
there is a penalty should it deviate from the Gaia-derived prior.

\subsection{Initial Model Fits}

Although the model produced a generally acceptable match to the observations, 
the initial runs of ELC did not yield satisfactory results in two ways. First, 
the best-fit primary star mass climbed as high as the model allowed 
($\sim$2 \Msun). 
This is inconsistent with the spectroscopically-determined temperature, 
photometric colors, and distance (see Section \ref{sec:2}). 
Based on the observed stellar temperature, metallicity, and radius from 
the Gaia parallax (and their uncertainties), and matching these with the 
Dartmouth stellar model isochrones \citep{Dotter2008}, we estimated that the mass of the primary star should be in the range 0.71 - 0.88 \Msun.
As a single-lined binary, the mass ratio is derived from the 
constraints placed by the three transits, not the radial velocity of the
secondary star. The transits perhaps provide only a very weak constraint
on the mass ratio, and so a highly uncertain primary mass is found.
But to strongly favor such a high mass with small uncertainty is implausible. 

The second disconcerting initial result was the estimated high mass of the 
planet. The planet's radius is well-determined by the light curve and the 
geometric and orbital constraints, and was found to be $\sim$3.6 \Rearth. 
It was therefore extremely unlikely that the $\sim$140 \Mearth \ mass that 
the models favored was accurate. For comparison, the empirical mass-radius 
relations from \citet{Lissauer2011} and \citet{Weiss2014} give 
$\sim$ 9-15 \Mearth.
A low-mass planet is also favored by the lack of variations seen in the
eclipse timing O--C diagram (see Section \ref{sec:planet}).

Looking closely at the residuals of the primary eclipse fits provided some 
insight (the secondary eclipses are very noisy and are all well-matched within their large uncertainties). 
There was correlated noise in the residuals, which we initially thought 
was due to the secondary star crossing a starspot on the primary star, as 
such occultations do result in large residuals and skew the mid-eclipse 
time (e.g.\ see Kepler-47 (\citealt{Orosz2012}; \citealt{Orosz2019}) and Kepler-453 \citep{Welsh2015}). However two lines of reasoning lead us to reject this 
hypothesis. First, the narrow eclipse profiles indicate that the eclipses 
are grazing, and this is confirmed by the initial model: the impact 
parameter was close 1.0. (In fact greater than 1.0 if the impact parameter 
is defined by the stellar radius only, not the sum of the radii.) 
The other line of reasoning comes from the lack 
of correlation between the eclipse-timing variations and the local slope 
of the light curve. Modulations in the light curve are due to starspots 
moving across the star's disk, and eclipses that cover the starspot skew 
the eclipse shape creating a deviation in the apparent mid-eclipse time 
\citep{Mazeh2015}. Such a correlation is seen on other 
circumbinary hosts e.g.\ Kepler-47, -453, and -1647 \citep{Kostov2016}. 
Though not impossible, it is unlikely that starspots reside so near the 
pole of the primary star assuming the star behaves like the Sun.
The lack of any correlation, despite Kepler-1661 certainly having starspots, 
supports the notion that starspots are not being eclipsed. Although no starspot crossing events were found,
looking more 
closely at the eclipse residuals revealed an interesting pattern: the 
model eclipses were in general too deep at early times and too shallow 
toward the end of the \kepler data. This could be the result of a change 
in the inclination of the binary caused by precession of its orbit. This 
in turn would favor a high-mass planet. This seems somewhat plausible, 
given that we observe the rapid precession of the planet (the transits 
grow significantly deeper over a span of less than a year). However, as 
described in the next section, we believe the primary eclipse depth change 
to be somewhat spurious, not a real consequence of a changing impact 
parameter.

% --------------------------------------------
\subsection{Eclipse Depth Variations: Cause and Effect}

The eclipses and transits are {\it relative} changes in the observed 
brightness of the system: the {\it Kepler} data that are modeled with ELC 
are normalized and detrended such that the out-of-eclipse flux is 1.0. For 
most cases this is fine, but for Kepler-1661 the eclipse depths are shallow 
and very sensitive to a change in impact parameter -- or an incorrect 
normalization that can occur if starspots are present.

The usual method to ``flatten'' the light curve outside of eclipses can 
introduce a bias in the eclipse depth if starspots are present. This is 
easy to visualize: suppose an eclipse of an immaculate star is 10\% deep. 
Now suppose a starspot blocks 50\% of the star's light. The 10\% deep 
eclipse will now appear to be 20\% deep in the normalized light curve. In 
Kepler-1661, the modulation that starspots create in the light curve are not 
only significant, but they are variable. It is this variability that is 
particularly troublesome. This 
changing starspot amplitude can induce an apparent change in the normalized 
eclipse depth. Although the starspot amplitudes are somewhat stochastic, 
there is a mild general trend towards larger amplitudes in the 
second half of the light curve, as shown in Figure \ref{fig:eclipse-depth} along with the measured primary eclipse depths.
This then has the effect of an overall 
increase in apparent eclipse depth. The measured eclipse depth variations, 
while very small ($\sim$ 2000 ppm), are a non-negligible fraction of the 
eclipses: $\sim$1\% RMS of the eclipse depth with a maximum change of $\sim$5\%. The photodynamical 
model attempts to fit this changing depth by changing the inclination via 
a precession in the orbit of the binary.

\begin{figure}[ht]
  \centering
    \includegraphics[width=1\textwidth]{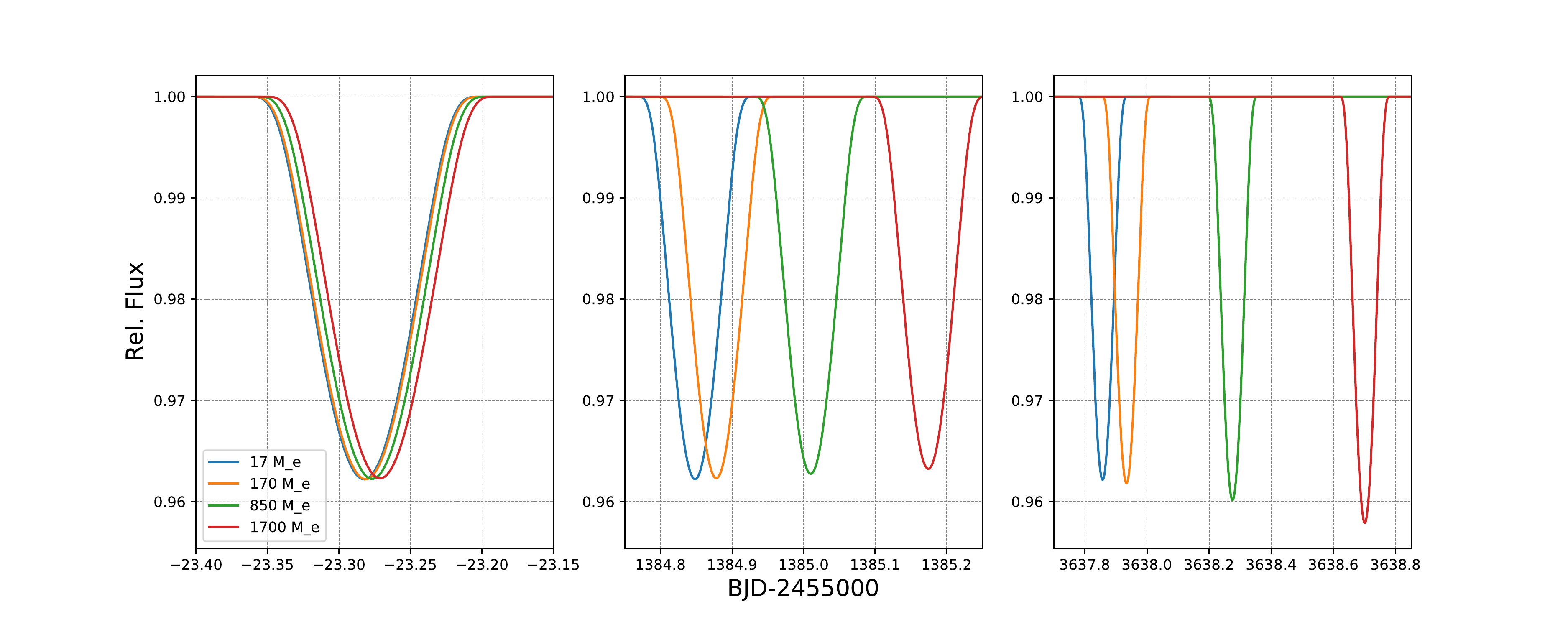}
  \caption{
Four different model primary eclipse profiles in the \textit{Kepler} bandpass are shown, spanning the duration
of the observations. The best-fit 17 \Mearth \ model is shown in blue. The other 
cases use identical parameters except for the planet mass: 170 \Mearth \ in orange, 
850 \Mearth \ in green, and 1700 \Mearth \ in red. 
In the middle panel the eclipse timing variations are easily seen as the shift
of the eclipses from the nominal case.
In the right-hand panel, the eclipse depth variations are now also easily noticeable.
There is also a more subtle change in eclipse width, although it is less pronounced than the depth variation.
  \label{fig:eclipse-depth-pred}}
\end{figure}

The mass of a CBP has traditionally been determined by
the eclipse timing variations (ETVs). The ETVs manifest themselves as the
divergence of the primary and secondary eclipse times in an O-C diagram, 
and also the ``ripples'' at the planet's orbital timescale that are superimposed on the long-term apsidal 
divergence.
However, a third observable signature is present:
the eclipse depth variations. To illustrate its effect, in
Figure \ref{fig:eclipse-depth-pred} we show model primary eclipse light
curves in the \textit{Kepler} bandpass that span the observations in this study. All parameters are
identical to the best-fit solution presented in Section \ref{sec:revised}, with the 
exception of the planet mass. Four cases are examined, with the planet
mass set to 17, 170, 850, and 1700 \Mearth.
At the start of the \kepler data the eclipses are identical in 
depth and width. Towards the end of the \kepler data the depths are still similar (though still measurable at \textit{Kepler} precision in the hundreds of ppm), 
however a significant eclipse timing variation can be seen for the larger 
planet masses. The mass of the planet simultaneously affects the eclipse timing, depth, and duration making them highly correlated. Pushing out to the epoch of the Mt.\ Laguna Observatory 
observation in 2019, the change in eclipse depth becomes readily apparent.
The change in depth is 
$\sim$29   ppm for a   17 \Mearth \ planet, 
$\sim$400  ppm for a  170 \Mearth \ planet, 
$\sim$2000 ppm for a  850 \Mearth \ planet, and 
$\sim$4400 ppm for a 1700 \Mearth \ (= 5.35 M$_{Jup}$) planet.
The point of this exercise is that eclipse depth variations depend on the
planet's mass, and inverting this, the planet's mass can be constrained by
the observed depth variations. However, this is true only if the depth 
variations are real changes in inclination of the binary, not created or biased by starspots; else, a spurious 
planet mass may be inferred. We believe this is the cause of the failure 
of our initial modeling. See Appendix \ref{app:one} for information on attempting to debias the eclipse depths.

% --------------------------------------------
\subsection{Revised Modeling and Results}\label{sec:revised}

\subsubsection{The Final Data Set and Isochrone Constraint}

Since the apparent eclipse depths vary depending on the presence of 
starspots, and the model is very sensitive to depth changes because of the 
grazing eclipse geometric configuration, we employed a technique that 
worked well for another CBP, Kepler-453 \citep{Welsh2015}. In that system, starspot modulations with a peak-to-peak variation 
of up to 1.5\% are present, and residuals of the fits to the eclipses 
clearly showed that the secondary star was sometimes eclipsing starspots 
on the primary star. These starspot-eclipse events skew the shape of the 
eclipse, resulting in an erroneous planet mass. (It is the eclipse timing 
variations that constrain the mass of the planet, and skewed eclipses 
produce spurious timing variations.) To mitigate the contamination caused 
by starspot eclipses in Kepler-453, only three clean primary eclipses were used. For the rest of the eclipses, only their eclipse times were used. Three eclipses were 
enough to the characterize the binary, and the eclipse times were 
statistically corrected for the starspot crossing bias by measuring, then 
removing, the correlation with the local light curve slope. For Kepler-1661, we 
employed the same technique, with one minor difference: No correlation of 
the eclipse timing variations and local slope was seen, and so no 
correction was applied. The three eclipses that were fit were selected 
at times when the starspot activity was low.

\begin{figure}[ht]
  \centering
    \includegraphics[width=1\textwidth]{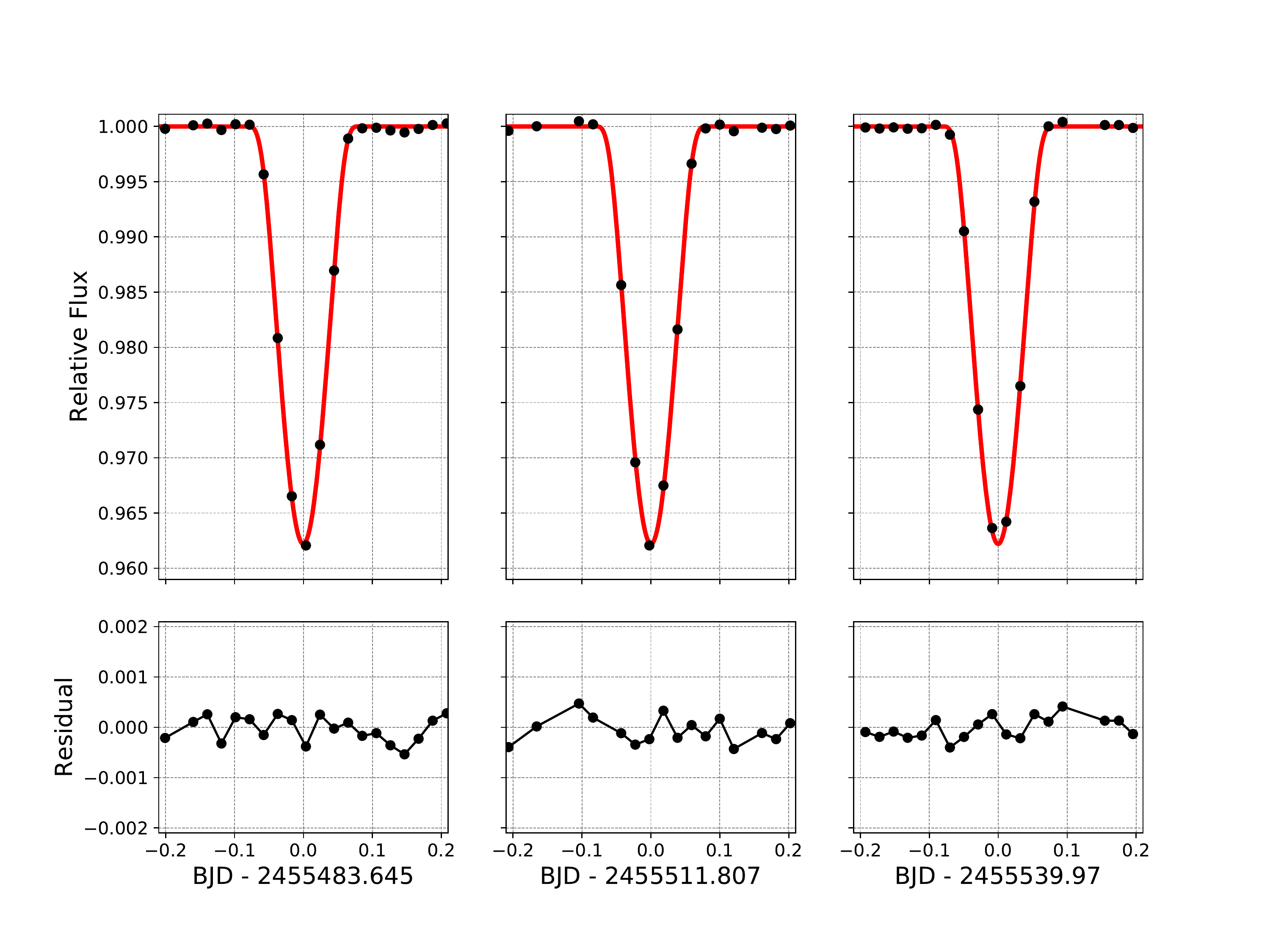}
  \caption{
The three primary eclipses that are fit with our photodynamical model.
The grazing eclipses create sharp, V-shaped eclipse profiles that are well-matched
by the model (shown in red).
  \label{fig:primss}}
\end{figure}

\begin{figure}[ht]
  \centering
    \includegraphics[width=1\textwidth]{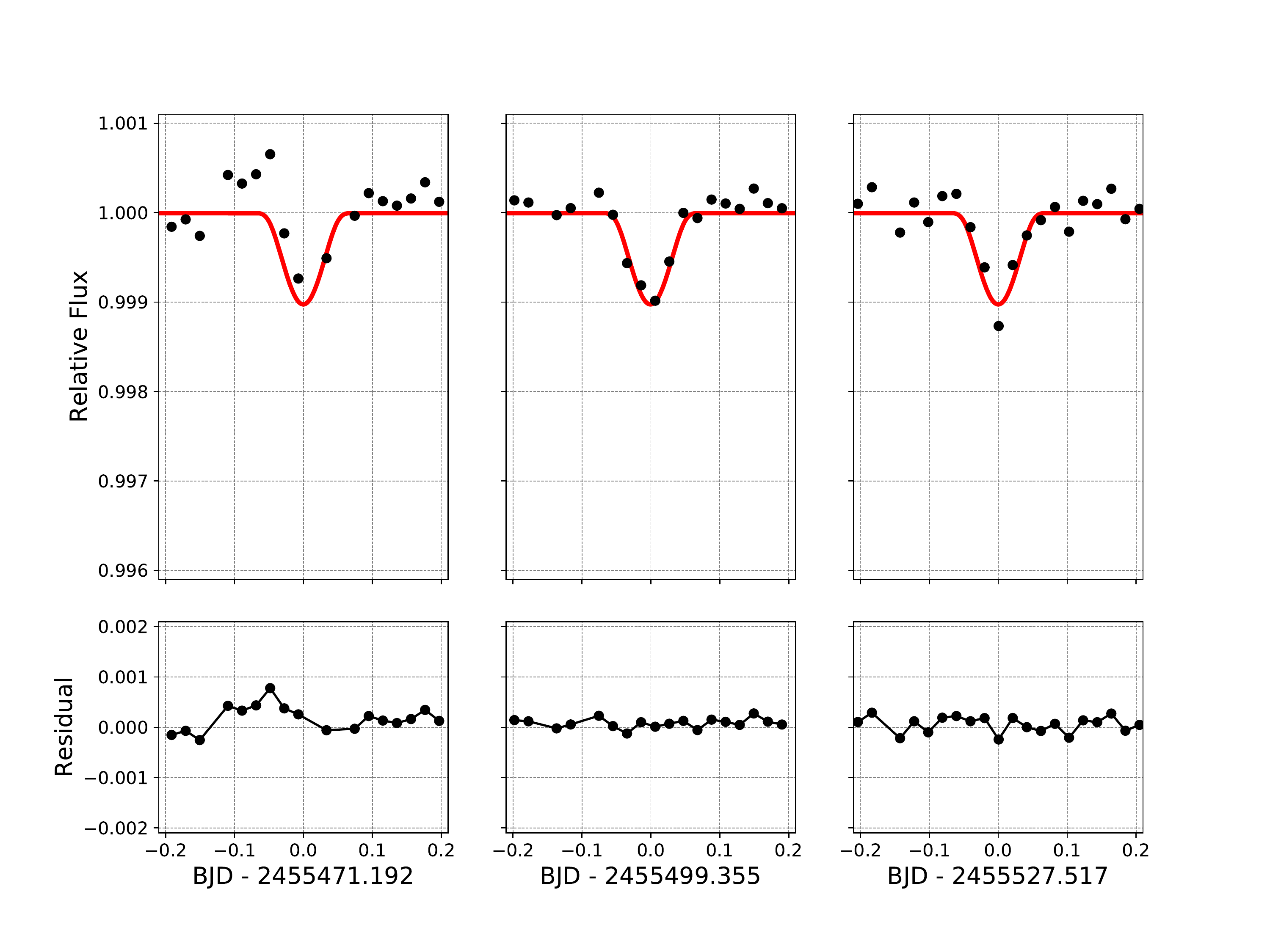}
  \caption{
Three examples of secondary eclipses and the best model fit.
The noisy, shallow eclipses are a limiting factor in determining the planet 
mass since the eclipse times cannot be measured with enough precision to allow
any meaningful constraint on the \textit{O-C} diagram. 
  \label{fig:secs}}
\end{figure}

Thus the final data set includes the three observed transits, 11 radial velocities, the Mt.\ Laguna 
primary eclipse, three {\it Kepler} primary eclipses, all observed \textit{Kepler} secondary eclipses, 
34 mid-eclipse times for the primary and 36 for the secondary, 
two windows of the {\it Kepler} light curve at times when 
transits {\it would} have occurred if the planet's impact parameter were 
not changing, and finally, one light curve window where a transit over the 
secondary could have occurred. In 
addition there are two additional data values, the temperature and primary 
radius measurements, for a grand total of 3457 data points. See Figure \ref{fig:primss} for the three primary eclipses used and Figure \ref{fig:secs} for the closest secondary eclipses.

Because this is a single--lined spectroscopic binary and the planet transits
do not constrain the mass ratio tightly enough to ensure a physically 
plausible stellar mass solution, we incorporate an additional feature into 
the ELC photodynamical model: a isochrone constraint. 
At each iteration, the trial solution's primary star mass and radius are 
compared to the PARSEC stellar isochrones \citep{Bressan2012}
that span ages from 1--10 Gyrs for a metallicity of -0.10. 
If for a given mass the radius is not within the range bracketed
by the isochrones, then a penalty is incurred. The penalty is treated as 
an addition to the $\chi^2$ value, and is computed as the 
square of the deviation of the radius from the 1 or 10 Gyr isochrone 
boundary, divided by 0.5\% of the radius value,
i.e., $((R - R_{\rm{boundary}}) / 0.005 R)^2$
This is akin to assuming a half-percent error bar on the radius.
The effect of this new feature is to steer the photodynamical solutions 
into a plausible region in the mass-radius plane.

\subsubsection{The System Parameters}

The ELC photodynamical model and the nested sampling and DE-MCMC techniques 
were able to satisfactorily fit the ``three-eclipses plus eclipse times'' 
data set. In particular, the mass of the primary star (0.84 $\pm$ 0.02 \Msun) 
and of the planet (17 $\pm$ 12 \Mearth) are very reasonable values. The best-fit 
$\chi^{2}$ is 4396 for 3457 degrees of freedom, or a reduced 
$\chi^{2}_{\nu}$ of 1.27. The DE-MCMC posteriors 
were generally Gaussian shaped with well-determined standard deviations. 
The exception to this were the limb darkening parameters. 
Our best-fit solution (lowest $\chi^{2}$) is presented in Table \ref{tab:fit} for the 
parameters fit by ELC. Table \ref{tab:deriv} list the system parameters, and Table \ref{tab:cart} gives 
the instantaneous velocities and positions of the three bodies at the 
reference epoch. The values allow an exact numerical integration and 
reproduction of our model, noting that the orbits are non-Keplerian 
and evolve rapidly with time. The orbital 
parameters listed in Tables \ref{tab:fit}, \ref{tab:deriv}, and \ref{tab:cart} are the instantaneous 
``osculating'' values valid only at the reference epoch and for short 
times thereafter. In Tables \ref{tab:fit} and \ref{tab:deriv}, note that $T_{conj}$ is the time of 
conjunction with the system's barycenter; for the binary this is 
approximately the time of mid-eclipse for the primary, but for the planet 
it need not be close to an actual transit time. Furthermore, this is the 
conjunction time based on the orbital elements at the reference epoch; 
since these evolve with time, the conjunction time will change as well.

% -------------------------------------------------------------------------------

\section{Discussion} \label{sec:discuss}

% --------------------------------------------
\subsection{The Binary}\label{sec:binary}
The binary consists of a K and M star in a 28.2-day, mildly eccentric 
($e_b$=0.112) orbit. The stars have masses of 0.84 $\pm$ 0.02 and 
0.262 $\pm$ 0.005 \Msun, and radii of 0.76 $\pm$ 0.01 and 
0.276 $\pm$ 0.006 \Rsun, consistent with being on the main sequence. 
Unlike some of the other circumbinary host stars, these stars do not 
have extremely precise mass and radius determinations, a consequence of 
having only three transits, all crossing the primary star at close to the 
same orbital phase.
The secondary star is much less luminous than the primary --
the ratio of secondary to primary bolometric luminosity is $\sim$3.3\%, 
and more specifically, it is only 1.1\% in the {\it Kepler} bandpass.

% ............
\subsubsection{Starspots and Stellar Rotation}\label{sec:rot}

\begin{figure}[ht]
  \centering
    \includegraphics[width=1\textwidth]{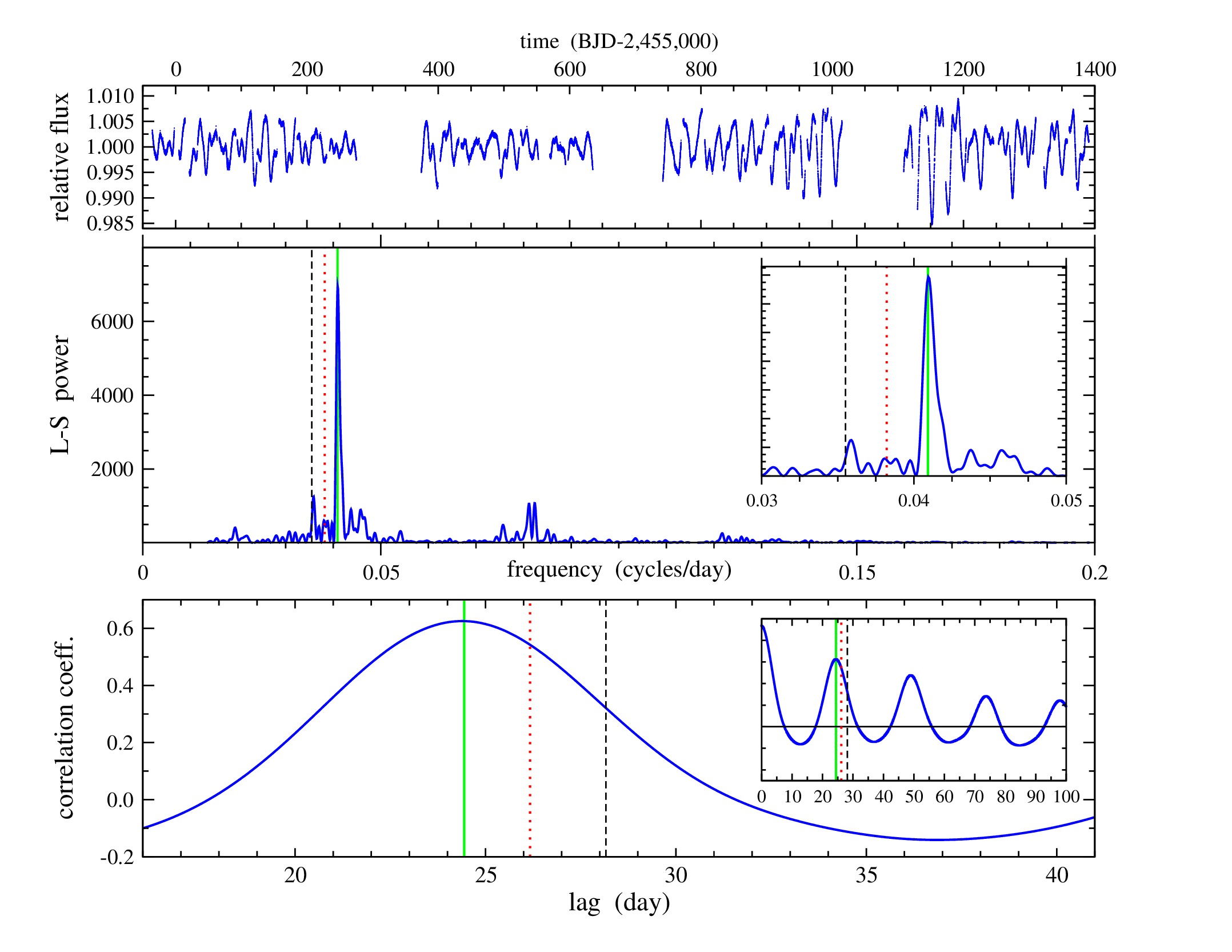}
  \caption{The upper panel shows the detrended and normalized light curve for Kepler-1661. 
Star-spot modulations are readily seen, and their amplitude is not constant over the 
4-years of {\it Kepler} data. The large $\sim$ 90-day gaps are due to the target falling 
on one of the failed CCD modules. The middle panels show the Lomb-Scargle power spectrum 
(the inset shows a zoomed-in version), with the black dashed line marking the orbital 
frequency, the red dotted line marking the pseudosynchronous frequency, and the green 
line marking the stellar spin frequency. The lower panels show the autocorrelation 
function (the inset shows a zoomed-out version). The orbital, pseudosynchronous, and spin 
periods are shown.
  \label{fig:spin}}
\end{figure}

The light curve of Kepler-1661 exhibits obvious quasi-periodic modulations 
that we interpret as being caused by starspots on the primary star. To 
measure the amplitude and period of the modulations, we first mildly 
detrend each Quarter to remove instrumental effects. The eclipses and 
transits were then removed, and any points with Data Quality flag greater 
than 16 were discarded, along with any obvious outliers and ramps due to 
the cooling of the photometer. A cubic polynomial was then used to detrend 
each Quarter. Both the SAP and PDC-MAP data were used in this analysis, 
and give consistent results. We also used the median of a 50-day wide 
sliding boxcar for the detrending, and it produced similar results. The 
SAP light curve is shown in the upper panel of Figure \ref{fig:spin}. We 
measure the peak-to-peak amplitude to be 2.5\% and the RMS variations to 
be 0.35\%. The starspot amplitude in Kepler-1661 is significantly larger than 
solar fluctuations ($\sim$ 0.1\%), implying a somewhat more active star. 
Note that the starspot modulation amplitude is not constant -- there are 
intervals when the starspots have very little effect on the light curve.

Assuming the quasi-sinusoidal modulations in the light curve are due to 
starspots on the primary star, we are able to measure the rotation period 
of the star. A discrete Fourier transform and a Lomb-Scargle periodogram 
were used to compute the power spectrum, after a 50\% split-cosine bell 
taper was applied to the detrended light curve. A strong spike at period 
24.43 days was found. Harmonics at two and three times the rotation 
frequency are seen, as well as a weaker peak is present at the binary 
orbital frequency (note that this is affected by a sidelobe of the window 
function of the time series).

Because Fourier techniques assume sinusoidal basis functions, they are not 
optimal for measuring periods of non-sinusoidal oscillations that change 
in both amplitude and phase. Hence we prefer to use the autocorrelation 
function (ACF) to measure the rotation period. However the standard ACF 
requires continuous data with uniform sampling, so we patched small gaps 
with a linear interpolation and patched larger gaps (e.g.\ when Kepler-1661 
was on a bad CCD module) with a random walk whose amplitude was scaled to 
match that of the light curve. We then created 100 realizations of the 
patched light curve and computed the ACF for each, then averaged. The 
result is show in the bottom panel of Figure \ref{fig:spin}. The peak of 
the ACF occurs at a period of 24.44 days. To estimate the uncertainty on 
the period, we also used the 2nd, 3rd, and 4th peaks in the ACF, dividing 
their periods by 2, 3, and 4. We then computed the weighted mean using the 
inverse of the correlation coefficient as the weight, and measured the 
standard deviation of the set. This is similar, but not identical to, the 
method described in \citet{2013MNRAS.432.1203M}. In particular, more weight 
is put on the first ACF peak. We repeated the above using the PDC-MAP 
light curve, and as a final sanity check, we also patched the light curve 
using pure white noise consistent with the RMS scatter of the light curve. 
All results were consistent with each other, and with the Fourier methods. 
We adopt as our final stellar rotation period estimate 24.44 $\pm$ 0.08 
days.

The measured stellar rotation period is less than the binary period, and 
more importantly, less than the 26.17 day pseudosynchronous period for an 
eccentric orbit \citep{Hut81} -- see Figure \ref{fig:spin}. However, this is 
not unexpected: For a 28-day period binary, the timescale for spin 
synchronization is over 25 Gyrs (and much, much longer for orbital 
circularization).

Using the measured rotation period and the estimate for the radius of the 
star, the expected $V_{rot} \sin{i}$ is 1.6 \kms, assuming the spin axis 
is perpendicular to the orbital plane and that the effects of any 
differential rotation are negligible. The observed $V_{rot} \sin{i}$ from 
the two highest signal-to-noise spectra is $\sim$ 2.5 $\pm$ 0.5 \kms, 
slightly higher than the estimate using the star's spin period.

% ............
\subsubsection{Comparison With Stellar Isochrones}
\begin{figure}[ht]
  \centering
    \includegraphics[width=1\textwidth]{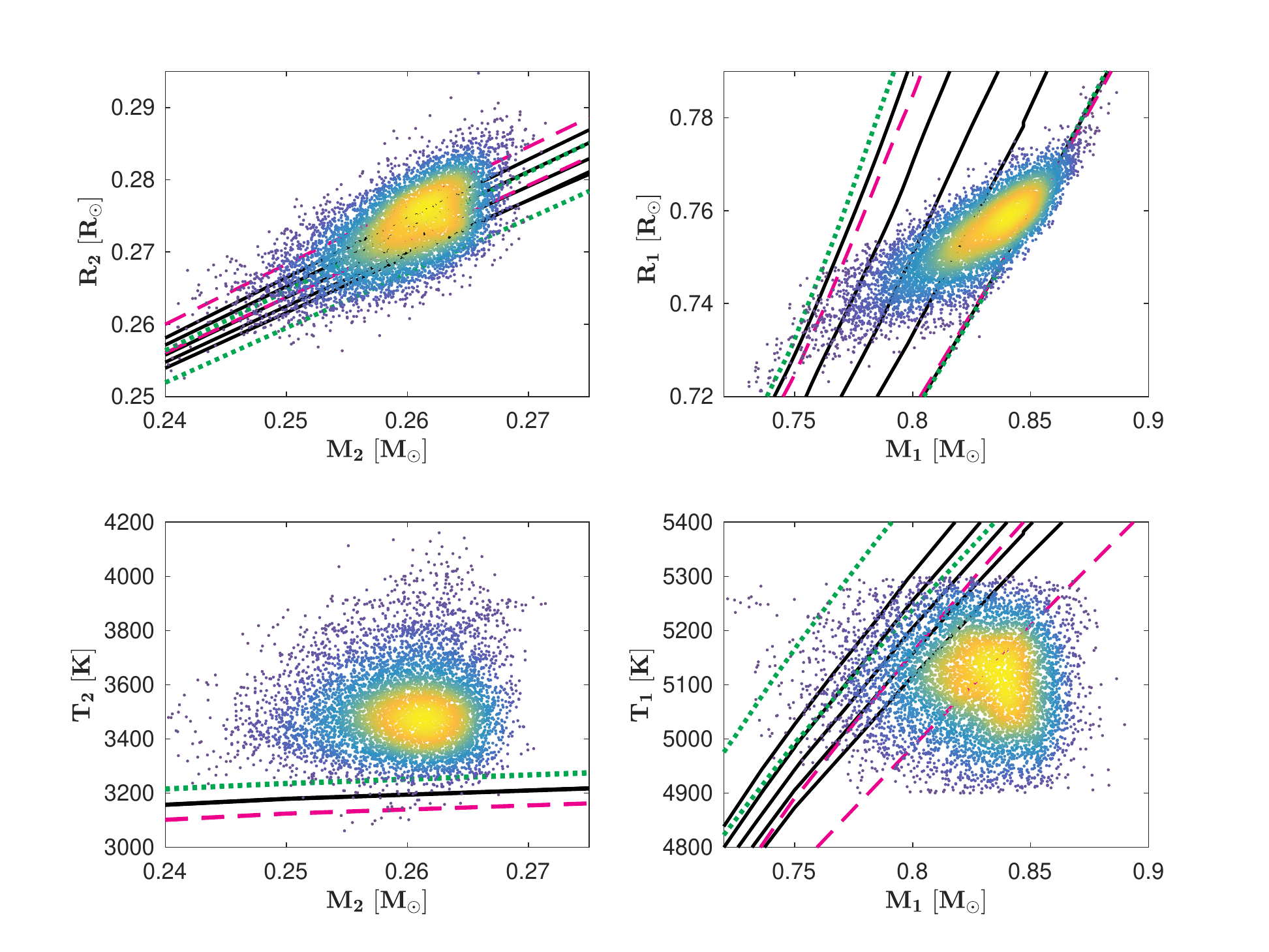}
  \caption{
The set of mass, radius, and temperatures from the MCMC posterior sample
for the secondary star (left panels) and primary star (right panels) are
plotted, along with the PARSEC isochrones \citep{Bressan2012}.
The solid black curves from bottom to top are the 1, 3, 5, 7 and 9 Gyr isochrones for the
nominal metallicity of $[{\rm Fe/H}]=-0.12$. 
The dashed magenta curves bracket the 1 to 9 Gyr isochrones 
for a metallicity of $-0.02$, and the dotted green curves are for a 
metallicity of $-0.22$. 
  \label{fig:isochrone}}
\end{figure}

In Figure \ref{fig:isochrone} we compare the MCMC posterior sample with the PARSEC 
isochrones \citep{Bressan2012}. 
The color of the points correspond to the density of the points in the 
figure, with blue being low density and yellow high density.
The primary star appears to be a relatively young star, 
$\sim$ 1--3 Gyrs, though there are some solutions in the posterior that 
extend up to 8 Gyrs (although with low probability). A young age is 
consistent with the starspot activity on the star.
A more solar-like metallicity is preferred than the nominal 
$[\rm{Fe/H}] = -0.12$,
and even higher-than-solar metalicities are favored if the stellar 
temperature is on the low end of its measured range.
The secondary star's radius is larger than expected; this is not unusual 
for stars of this mass (e.g.\ see the review by
\citet{Torres2010}). The temperature is also higher 
than expected, and this is somewhat atypical, though there is large 
uncertainty in the temperature.

% ---------------------------------------------------
\subsection{The Planet}

% ............
\subsubsection{Planet Characteristics}\label{sec:planet}

\begin{figure}[ht]
  \centering
    \includegraphics[width=1\textwidth]{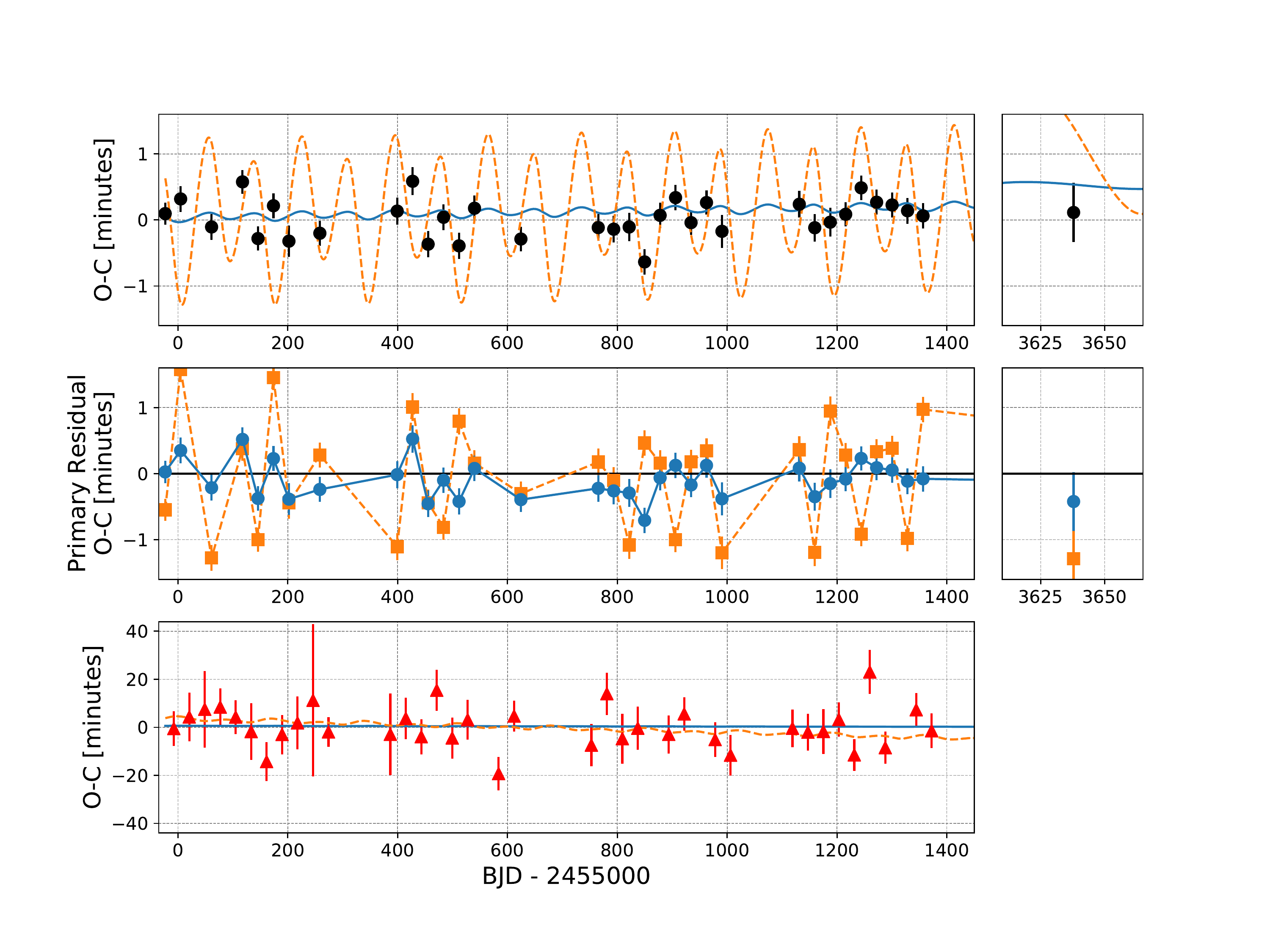}
  \caption{
The observed-minus-calculated \textit{O-C}
mid-eclipse times for the \kepler primary eclipses (top panel) 
and secondary eclipses (bottom panel) after the best-fit, 
common linear ephemeris has been subtracted. 
The upper right-hand panel corresponds to the MLO primary eclipse.
The blue curve is the best-fit 17 \Mearth \ model prediction.
The dotted orange curve is the best-fit model for a 1 \Mjup \ planet. The middle panel shows the residual primary eclipse times against the models shown in the upper panel, with the blue-circles being the residuals of the nominal 17 \Mearth \ model and orange squares for the residuals of the 1 \Mjup \ model for the mass of the planet.
While the primary eclipse times are noisy and are not particularly well matched 
by the nominal model, the Jupiter-mass model is much worse 
($\chi^2$ of 79 versus 594 with 25 fitting parameters and 34 data points).
  \label{fig:oc}}
\end{figure}

The {\it Kepler} light curve contains three transits across the primary star, 
substantially fewer than the eight that could potentially have been detected 
given the 175~d period of the planet. However, three full Quarters are of data 
are missing (plus Quarter 17) because of the failed CCD module, and the planet's 
impact parameter was greater than 1.0 prior to the third year of observations. In 
addition, a transit that could have been detected in Quarter 16 fell in a small 
data gap. No transits over the secondary are detected, although the non-detection of any 
such transits is fully consistent with the secondary star being much fainter 
than the primary. The three transits do provide enough information to 
characterize the planet fairly well, though much of the uncertainty is propagated 
from uncertainty in the binary star parameters. The planet's radius (3.87 $\pm$ 
0.06 \Rearth) is well-determined and similar to Neptune's (3.88 \Rearth). The 
mass, however, is much less well-determined: 17 $\pm$ 12 \Mearth. This is a 
consequence of the weak constraint placed by the eclipse timing variations. The 
eclipse timing variations, expressed in a common-period O--C diagram is shown in 
Figure \ref{fig:oc}. The secondary times are very noisy due to the shallow 
secondary eclipses, and this prevents any useful mass constraint based on an 
induced apsidal motion of the binary. The main mass constraint is therefore based 
on the ``ripples'' in the O--C, caused by the dynamical 
perturbation of the binary by the planet (\citealt{Borkovits2011}; \citealt{Borkovits2015}). 
While small, the effect the planet has on the binary still dominates over the
general relativistic precession (which accounts for 17\% of the precession) and
the classical apsidal motion due to tidal interaction (less than 1\%).
Given the small amplitude of the O--C variations, it is perhaps more correct 
to say that it is the {\it lack} of eclipse timing variations that provides 
an upper limit constraint on the planet's mass.
This is illustrated in Figure \ref{fig:oc}  where the orange curve shows the 
expected O-C variations for a planet of 1 \Mjup. 
The variations from a planet of this mass is larger than the observed 
variations, thus the planet is of lower mass. The blue curve shows the 
photodynamical model best-fit variations, and while not a particularly good 
match, it is much more consistent with the amplitude of the variations. Despite 
the low-precision mass determination, the conclusion is robust: the circumbinary 
object in Kepler-1661 is substellar. With three transits that match in detail the 
times, depths, and durations expected of a circumbinary object, the candidate's 
planethood is established.

% ............
\subsubsection{Orbital Characteristics}

The planet's orbit is mildly eccentric ($e_p\approx 0.057$) and resides nearly 
co-planar ($\Delta i \sim 1^\circ$) with the binary orbital plane, which is 
consistent with the orbital properties of all the known transiting Kepler 
CBPs \citep{Li2016}.  The planetary orbital period ($P_p = 175.06 
\pm 0.06$) is $\sim$6.2 times the binary orbital period.  Using the the stability 
criteria from \citet{Holman1999}, we find the ratio $P_p/P_{crit}$ =1.381 
indicating that the planet is on the stable side of the so-called stability limit.  
The period ratio between the innermost planet and the binary is an interesting characteristic of many \textit{Kepler} CBPs, and \cite{Welsh2018} note this ratio is close to unity for many CBP systems. In terms of the planetary 
semimajor axis ratio $a_p/a_{crit}$ (=1.281), the planet also appears to be near 
the stability limit.  However, the \citet{Holman1999} analysis produces a 
stability formula that is averaged over several parameters, where the stability 
limit can be over- or under-estimated when compared to n-body simulations of 
specific systems.  In this case, the \cite{Holman1999} criterion overestimates the critical 
period $P_{crit}$ for stability and hence the ratio $P_p/P_{crit}$ is larger 
($\sim$1.434) if more sophisticated analyses are used 
\citep{Lam2018,Quarles2018}.  These analyses are generally applicable for planets 
on near circular and co-planar orbits, but \cite{Quarles2018} provides an 
empirical relationship for the maximum eccentricity for a planet as a function of its semimajor axis ratio ($e_{max} 
\approx 0.2$ for Kepler-1661) before it becomes unstable due to the overlap of N:1 mean motion 
resonances with the binary \citep{Mudryk2006,Sutherland2019}.  The best-fit 
planetary eccentricity could triple and remain below the threshold for 
eccentricity, which further enhances the evidence for a stable orbit.

Another definition for a CBP to be at the stability limit is to check if another planet of equal mass is allowed in between the known planet's orbit and $a_{crit}$.
We use a planet-packing formalism that uses the dynamical 
spacing $\beta$ between planets \citep{Chambers1996,Kratter2014,Quarles2018}, 
where $\beta \geq 7$ indicates that an equal-mass nearby planet with semimajor 
axis $a_{crit}$ would also be stable.  Using the stability fitting formula from 
\cite{Holman1999}, we find $\beta \approx 7.28$ and demonstrates that although this planet 
is near the stability limit, it still allows for a stable interior planet.  Furthermore, we 
can use the more recent approaches and find that the spacing of the planet 
relative to the stability limit increases, where $\beta \approx 9.18$ using the 
machine learning method \citep{Lam2018} and $\beta \approx 8.21$ using the grid 
interpolation method \citep{Quarles2018}.
    
\begin{figure}[ht]
  \centering
    \includegraphics[width=1\textwidth]{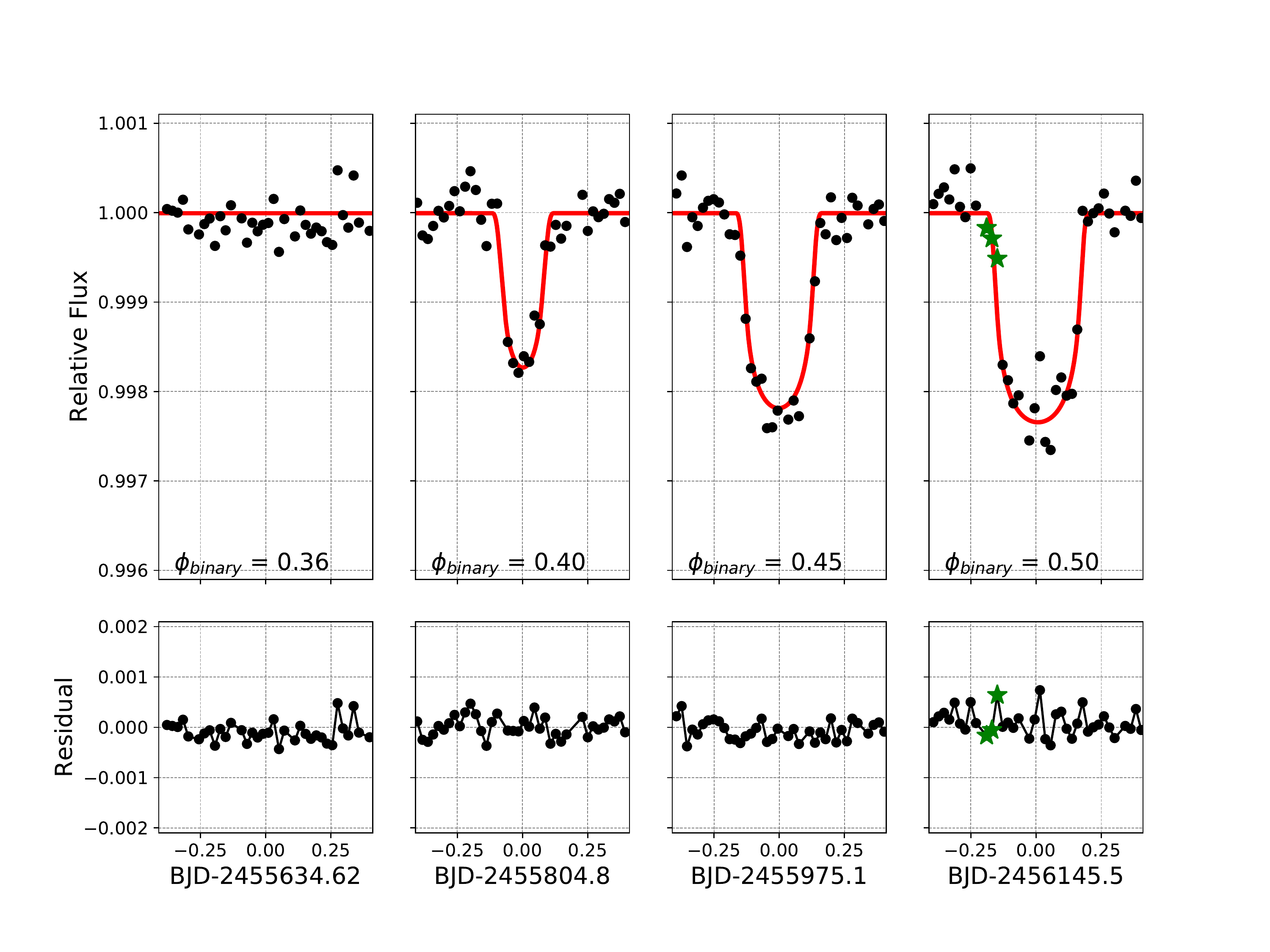}
  \caption{
Transits of the planet across the primary star, and the best-fit model.
The changing depths and widths are a classic signature of a circumbinary object. 
In general, the depth and width changes are due to the varying relative position 
and velocities of the star and planet at the times of transit. However for 
Kepler-1661, the orbital phase of the binary during the transits are similar, 
so the changes in the transit profiles are mainly due to the changing 
impact parameter caused by the precession of the planet's orbit.
The orbital phase of the binary star is given in the upper panels.
The first panel shows no transit, though this is where one would expect a
transit to occur if the planet's orbit did not precess so much that the absolute value of the impact
parameter is greater than one.
The three green ``star'' points in the rightmost panel lie near a rejected 
observation, and they have had their uncertainties boosted by a factor of 10.
Lower panel: Residuals (data minus model) of the fits.
  \label{fig:trans}}
\end{figure}

\begin{figure}[ht]
  \centering
    \includegraphics[width=1\textwidth]{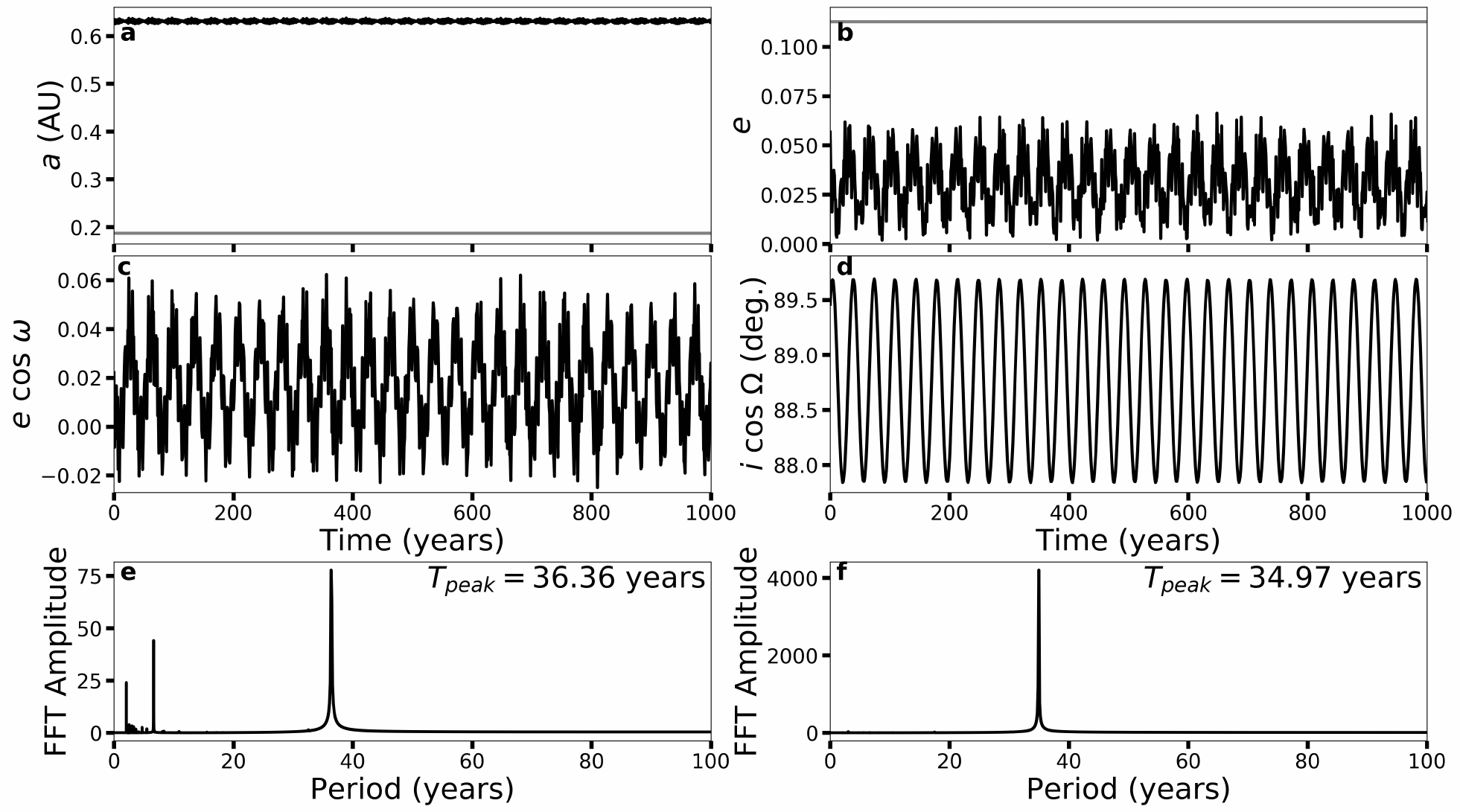}
  \caption{
Orbital evolution for 1,000 years starting from BJD = 2454960 of the semimajor axis (panel (a)), and the 
eccentricity (panel (b)), for the planet and binary (black curve and gray curve, 
respectively).
The evolution for the x-component in the (c) eccentricity $e\cos{\omega}$ and (d)
inclination $i\cos{\Omega}$ vector is given, along with a periodogram showing the periods of (e) apsidal precession from $e\cos{\omega}$ and (f) nodal precession from $i\cos{\Omega}$.
  \label{fig:longterm}}
\end{figure}

\begin{figure}[ht]
  \centering
    \includegraphics[width=1\textwidth]{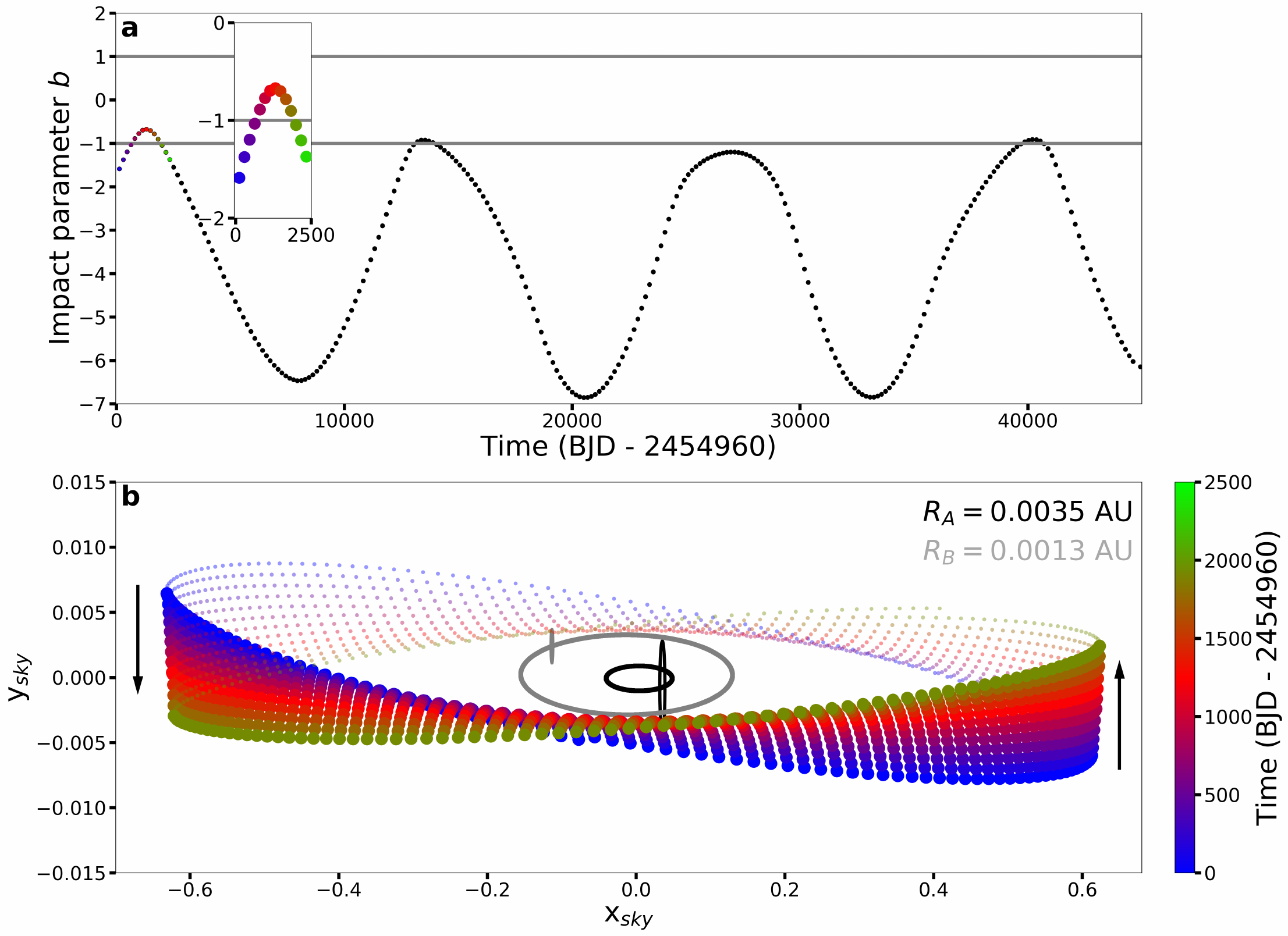}
  \caption{
The effect of nodal precession on the impact parameter $b$ is shown in the upper
panel (a), over the span of 45,000 days. Only for those conjunctions with $|b| < 1$ do transits occur, as denoted by the horizontal gray lines. The long-term oscillations in the impact 
parameter limit the transit-ability of Kepler-1661, where some parts of the precession 
cycle prohibit transits (e.g.\ near time $\sim$27,000 days).  The inset shows a 
zoomed-in view of the first 2,500 days, where only 7 transits are  possible.  
Panel (b) shows the plane of the sky alignment of the planet and binary and
illustrates how the orbit of the planet tilts due to nodal precession.
The points are color-coded with respect to the time in days, which spans
a range of 2500 days, or about 20\% of a precession cycle. The 
cross-section of stellar disk for star A (black) and star B (gray) are shown as 
thin hoops, stretched vertically because of the very different y-axis scale. The other two ellipses show their orbits.
The transparency of the points indicate the z-component of the planetary orbit 
on the sky plane, where the smaller, faint, semi-transparent points lie into the page 
(behind the barycenter).
  \label{fig:impact}}
\end{figure}

We dynamically integrated the system orbits using the IAS15 integrator in the REBOUND code 
\citep{Rein2012,Rein2015}, starting with the parameters listed in Tables 
3 and 4.  Our 100,000 year simulation revealed small oscillations in the 
planetary semimajor axis, eccentricity, and inclination, where the maximum 
eccentricity and inclination was 0.066 and 1.57$^\circ$, respectively.  Figure \ref{fig:longterm} 
shows the evolution of the planetary (black) and binary (gray) semimajor axis 
and eccentricity for the first 1,000 years.  We also show the evolution for the 
x-component (middle panels) of the planetary eccentricity $e\cos{\omega}$ and inclination 
vector $i\cos{\Omega}$ along with the corresponding periodogram (bottom panels) that 
illustrates the planetary apsidal and nodal precession periods.  Although the two 
precession periods are similar in the case of Kepler-1661, the 35-year nodal 
precession period is more relevant for transits.  A result of the rapid 
precession of the planet's orbit is clearly seen in Figure \ref{fig:trans}, where the transit 
depth and width increase on subsequent transits.  Such a rapid precession is not 
unusual for CBPs, where Kepler-413 has a remarkably short 11-year 
precession timescale \citep{Kostov2014}.  A consequence of the nodal precession 
is that the orbital inclination, and hence the impact parameter, b, is 
continuously changing.  Figure \ref{fig:impact} shows the variation in impact parameter, ranging from -6.8 to -0.67,
over the course of 45,000 days or $\sim$3.5 precession 
periods. Interestingly, as the planet's orbit precesses, the plane of the 
planet tilts up from below the binary plane to cross the primary, but it never 
gets as high as the equator before precessing back down off the star. The 
planet can only transit the host star if $|b|<1$, and as can be seen in Figure 
\ref{fig:impact}, this is a small part of the curve. The fraction of time a precession 
cycle that the planet is sufficiently aligned with our line of sight to enable 
transits is only $\sim$7\% on average, noting that one out of every four 
precession cycles results in no observable transits.

In Figure \ref{fig:impact} we also show how the planet's orbital plane tilts on the plane of 
the sky over the first 2500 days (color-coded), which further demonstrates the 
rarity of alignments that allow for transits.  At the current time, the planet is 
no longer transiting. The next cycle of transits is expected to start in 2045.
See Table \ref{tab:trans} for the predicted times, impact parameter, and 
durations of future transits. These are for the best-fit model, but of course 
there is a spread of acceptable solutions, so there is a range of values for the 
items in the table.

% ............
\subsubsection{Habitable Zone}

In general, the {\it Kepler} CBPs that are located near the 
orbital dynamical instability limit are often near their classical stellar-heated habitable zones. Kepler-1661 
is no exception. Although its mass and radius suggest a ``warm Neptune'' planet 
that is not conducive to life as we know it, it is still interesting to 
estimate the amount of radiant energy the planet receives and compare this 
with the habitable zone.

\begin{figure}[ht]
  \centering
    \includegraphics[width=1\textwidth]{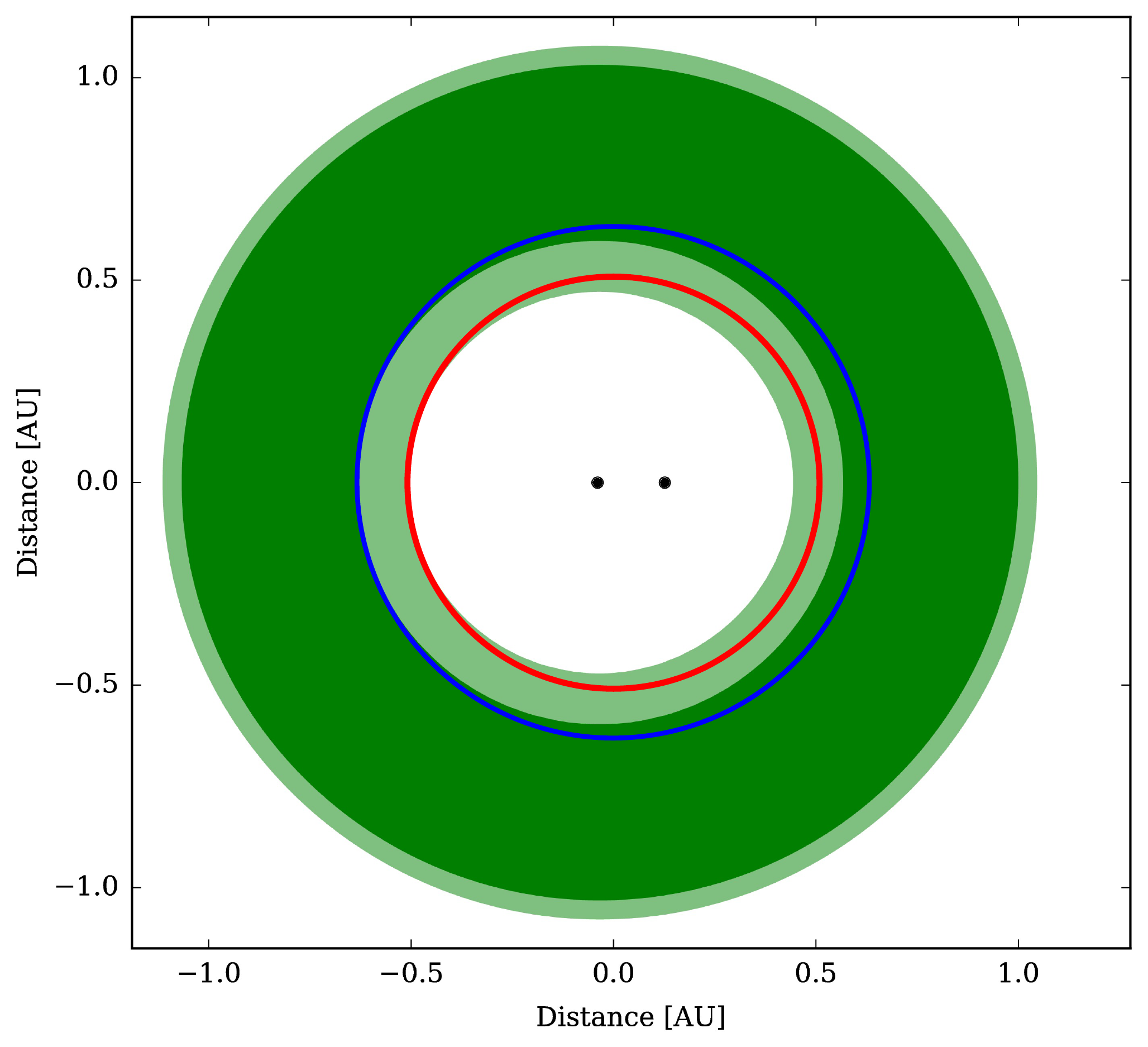}
  \caption{
Face-on view of the Kepler-1661 system, showing the planet's 
orbit (in blue) relative to the binary and the habitable zone.
The dark green region corresponds to the narrow (conservative) habitable zone, 
and the light green corresponds to the nominal (extended) habitable zone 
as defined by \citet{Kopparapu_2013a} and \citet{Kopparapu_2013b}. The critical radius for stability is 
shown in red.
  \label{fig:hz}}
\end{figure}

The K-star primary dominates the energy output of the stars, allowing a 
first-order approximation to the incident stellar flux on the planet's 
atmosphere to be easily computed. 
The orbit-averaged insolation at the reference epoch 
is 0.88 $S_{\earth}$ where 
$S_{\earth}$ is the Sun-Earth insolation (equal to 1367 $\rm{W\ 
m^{-2}}$). This insolation is less than the conservative ``moist greenhouse'' 
upper limit of 0.961 $S_{\earth}$  for the hotter inner edge of the habitable 
zone and well within the more optimistic runaway 
greenhouse or recent Venus limits (see \citet{Kopparapu_2013a}, \citet{Kopparapu_2013b}, and the 2014 on-line updated 
coefficients\footnote{http://depts.washington.edu/naivpl/content/hz-calculator}). Assuming a Bond albedo of 0.34 and that the planet re-emits the 
absorbed radiation over a full sphere, the planet's equilibrium temperature 
$T_{\rm{eq}}$ is $\sim$ 243 K. A face-on view of the Kepler-1661 system is shown 
in Figure \ref{fig:hz}, created with the web-based software Multiple Star HZ 
calculator\footnote{\tt http://astro.twam.info/hz/} described in \citet{Mueller_2014}. The darker green region corresponds to the conservative 
habitable zone and the lighter green corresponds to the optimistic habitable 
zone. The red circle marks the (in)stability radius based on 
the \citet{Holman1999} formula.

\begin{figure}[ht]
  \centering
    \includegraphics[width=1\textwidth]{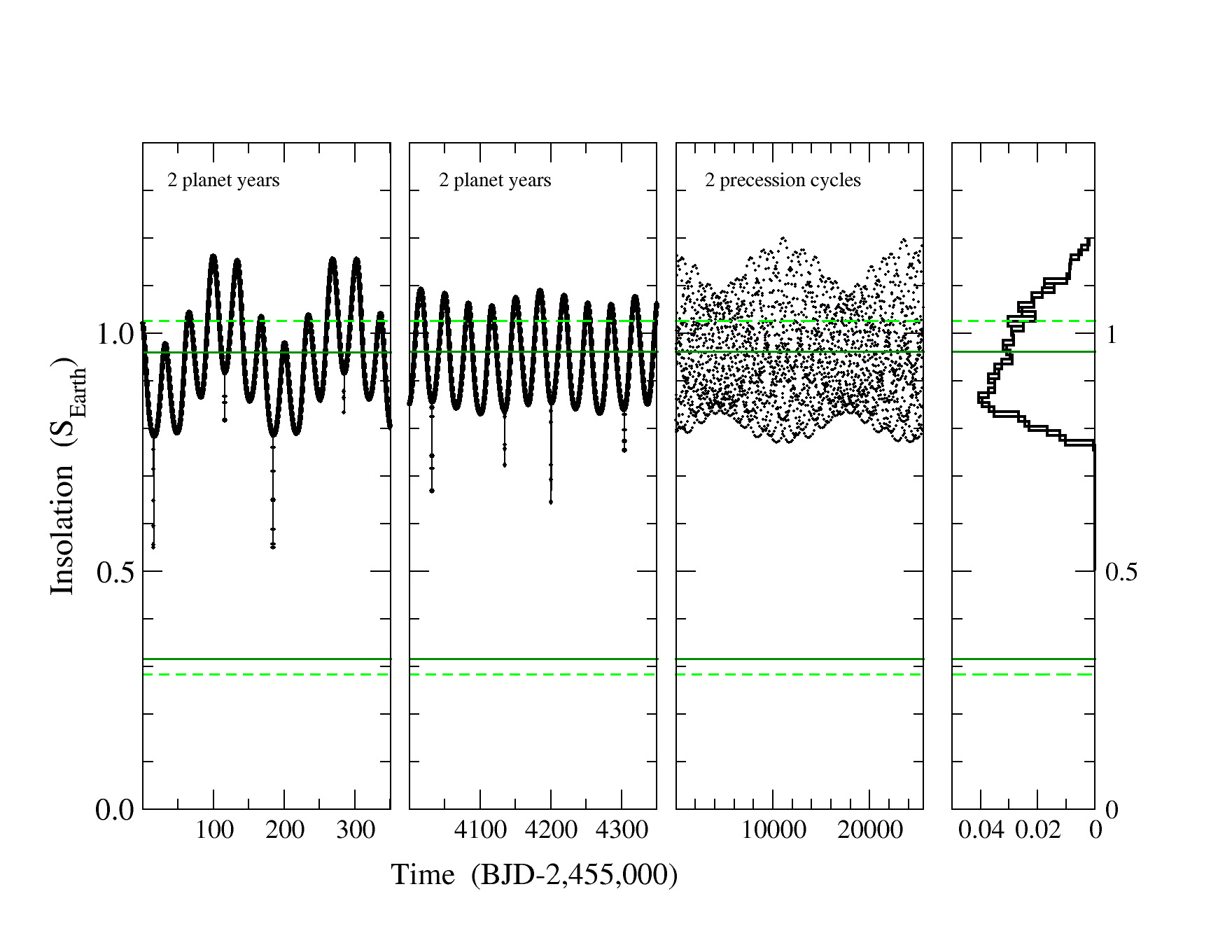}
  \caption{
The insolation incident at the top of the atmosphere for Kepler-1661, in units
of the Sun-Earth insolation.
{\it Left panels:} The fluctuations in the insolation over a period
of two planet orbits (2 $\times$ 175.4 days) during the {\it Kepler} 
epoch. The dashed green lines mark the boundaries of the habitable zone
as defined by \citet{Kopparapu_2013a} and \citet{Kopparapu_2013b}.
The sharp downward spikes are due to the  stellar eclipses as seen from the planet.
{\it Middle-left panel:} The insolation at an epoch where the fluctuations are less 
extreme.
{\it Middle-right panel:} The distribution of the insolation over two apsidal precession 
cycles of the planet (equal to $\sim$70 years).
{\it Right panel:} A histogram of the long-term distribution of the insolation.
  \label{fig:insolation}}
\end{figure}

Integrating the equations of motion for the three bodies allows us to compute 
the exact instantaneous insolation, and enable us to follow it though a 
precession cycle. Figure \ref{fig:insolation} shows the insolation from both stars as a function 
of time. The shortest timescale variation variation ($\sim$28 days) is caused 
by the orbital motion of the primary star. Variation due to the planet's eccentric 
orbit are present at a timescale of the orbital period of the planet.
On a much longer timescale are fluctuations caused by the 
precession of the orbits ($\sim$35 years). The precession causes the peak-to-peak
fluctuations in insolation over the course of the planet's year to vary cyclically.
The average insolation over the 
full precession cycle is $<$S$>$=0.947, with an RMS fluctuation of 10.4\%.
The median and the mode are slightly lower, at 0.936 and 0.865, respectively.
These long-term averages are just within the conservative HZ limit (set by the 
moist greenhouse criteria), though excursions above that limit, and even above 
the runaway greenhouse limit, are present for a substantial amount of time -- see
the histogram in the last panel of Figure \ref{fig:insolation}.

% ----------------------------------------------------------------------------------
\section{Conclusion}

In this study, we report the discovery of a transiting circumbinary planet of 
approximately Neptune mass and radius in a nearly coplanar orbit around a 
K + M eclipsing binary. The planet orbits near the critical stability radius 
and is, on average, inside the habitable zone.
The host binary is a single-lined spectroscopic binary but the planet transits 
can, in theory, provide enough information to determine the absolute masses, 
radii, and geometry of the system.
Unfortunately, in Kepler-1661 only three transits across the primary star were observed and they all
occur near the same binary phase, and thus provide only limited constraints
on the primary star's position and velocity. No transits were observed across the secondary. However, by steering the 
photodynamical modelling solutions to agree with stellar isochrone models, 
enough information is available to determine a full set of system parameters. 
Care must be taken in the modelling because the eclipses of the stars are 
grazing (high impact parameter), making them particularly sensitive to the 
orbital inclination. The planet causes the binary's orbit to precess, which 
causes the inclination to change, which in turn causes the eclipse depths to 
change. But starspots create modulations in the light curve, and the standard 
procedure for detrending and normalizing \kepler data can also result in 
(spurious) changes in eclipse depth.

A ground-based observation of a primary eclipse was obtained in 2019 and was 
extremely valuable because it more than doubled the temporal baseline of the 
time series. This allowed a much better determination of the effect the planet 
has on the binary. The precession period is 35 years, and the planet spends
only $\sim$7\% of the time in a configuration where transits are detectable from
our line of sight. The transits which began in 2011, ended in 2014, and the 
next cycle of observable transits is not expected to start until 2045. 
Thus no transits are expected during the \textit{TESS} Primary Mission.

% ------------------------------------------------------------------------------------
\acknowledgments
We thank the anonymous reviewer for suggestions and comments that have improved this paper. 
We gratefully acknowledge support from the National Science Foundation via 
award AST-1617004, and we are are also deeply grateful to John Hood, Jr.\ 
for his generous support of exoplanet research at SDSU.
We acknowledge years of fruitful discussion with members of the Kepler 
TTV/Multi-Planet Working Group and Eclipsing Binaries Working Group.
In particular we would like to thank Josh Carter for initial work on this 
CBP system.
We also thank Andre Prsa, Kelly Hambleton, Kyle Conroy (Villanova Univ.), 
and Tara Fetherolf and Trevor Greg (SDSU undergraduates at the time)
for their assistance in acquiring the KPNO spectra.
\textit{Kepler} was competitively selected as the 10th mission of the Discovery
Program. Funding for this mission is provided by NASA, Science Mission
Directorate.
The Hobby-Eberly Telescope (HET) is a joint project of the University of Texas at Austin, the Pennsylvania State University, Ludwig-Maximilians-Universit\"at M\"unchen, and Georg-August-Universit\"at G\"ottingen. The HET is named in honor of its principal benefactors, William P. Hobby and Robert E. Eberly.
This research has made use of the Exoplanet Follow-up Observation Program website
and the NASA Exoplanet Archive, which are operated by the California Institute of 
Technology, under contract with the National Aeronautics and Space Administration 
under the Exoplanet Exploration Program.
This work is also based in part on observations at Kitt Peak 
National Observatory, National Optical Astronomy Observatory, which is operated 
by the Association of Universities for Research in Astronomy (AURA) under a 
cooperative  agreement with the National Science Foundation. 
Simulations in this paper made use of the REBOUND code which is freely available 
at \url{http://github.com/hannorein/rebound}.

% ------------------------------------------------------------------------------------
\facilities{Kepler, Smith (Tull spectrograph), HET (HRS spectrograph), Mayall (Echelle 
spectrograph), MLO:1m}

\software{AstroImageJ \citep{Collins2017}}

% ------------------------------------------------------------------------------------

% ------------------------------------------------------------------------------------
% ------------------------------------------------------------------------------------

% ++++++++++++++++++++++++++++++++++
% FIGURE 1
% ++++++++++++++++++++++++++++++++++

% ++++++++++++++++++++++++++++++++++
% FIGURE 2
% ++++++++++++++++++++++++++++++++++

% ++++++++++++++++++++++++++++++++++
% FIGURE 3
% ++++++++++++++++++++++++++++++++++

% ++++++++++++++++++++++++++++++++++
% FIGURE 4
% ++++++++++++++++++++++++++++++++++

% ++++++++++++++++++++++++++++++++++
% FIGURE 5
\begin{figure}[ht]
%\epsscale{0.50}
  \centering
    \includegraphics[width=0.8 \textwidth]{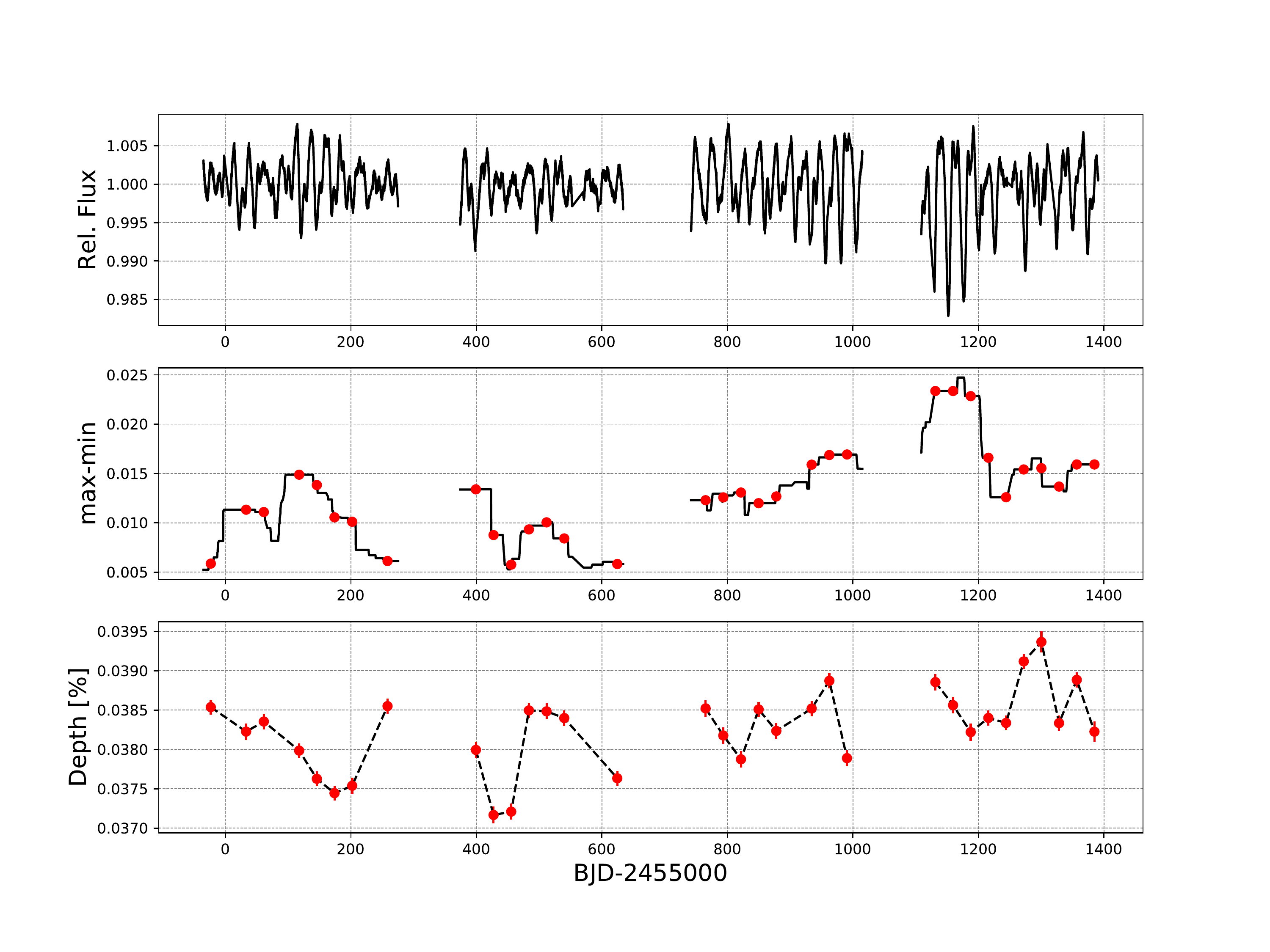}
  \caption{
{\it Upper panel:} The normalized \kepler light curve sans eclipses, highlighting the 
starspot modulations.
{\it Middle panel:} The peak-to-peak amplitude of the starspot modulation over a 50-day 
wide sliding window. The red dots mark the location of observed primary eclipses.
{\it Bottom panel:} The measured primary eclipse depths. Note the slight overall upward 
trend over the course of the observations, indicating an increase in eclipse depth.
  \label{fig:eclipse-depth}}
\end{figure}
% ++++++++++++++++++++++++++++++++++

% ++++++++++++++++++++++++++++++++++
% FIGURE 6
% ++++++++++++++++++++++++++++++++++

% ++++++++++++++++++++++++++++++++++
% FIGURE 7
% ++++++++++++++++++++++++++++++++++

% ++++++++++++++++++++++++++++++++++
% FIGURE 8
% ++++++++++++++++++++++++++++++++++

% ++++++++++++++++++++++++++++++++++
% FIGURE 9
% ++++++++++++++++++++++++++++++++++

% ++++++++++++++++++++++++++++++++++
% FIGURE 10
% ++++++++++++++++++++++++++++++++++

% ++++++++++++++++++++++++++++++++++
% FIGURE 11
% ++++++++++++++++++++++++++++++++++

% ++++++++++++++++++++++++++++++++++
% FIGURE 12
% ++++++++++++++++++++++++++++++++++

% ++++++++++++++++++++++++++++++++++
% FIGURE 13
% ++++++++++++++++++++++++++++++++++

% ++++++++++++++++++++++++++++++++++
% FIGURE 14
% ++++++++++++++++++++++++++++++++++

% ++++++++++++++++++++++++++++++++++
% FIGURE 15
% ++++++++++++++++++++++++++++++++++

\clearpage
% ------------------------------------------------------------------------------------
% . . . . . . . . . . . . . . . . . . . . . . . . . . . . . . . . . . . . . . 

\begin{table}[ht]
	\centering
	\caption{Kepler-1661 Radial Velocity Measurements}
	\label{tab:rvs}
	\begin{tabular}{llll}
		\hline
        \hline
	Date (BJD) & RV (km s$^{-1}$) &  $\sigma$ (km s$^{-1}$) & Instrument \\ \hline 
	2456017.921467 &  17.42    &  0.07 &  HET \\
        2456022.921967 &  -0.46    &  0.04 &  HET \\
        2456039.871467 &  30.20    &  0.06 &  HET \\
        2456138.837280 &  -3.34    &  0.05 &  HET \\
        2456230.586181 &  10.44    &  0.06 &  HET \\
        2456083.794858 &  -2.59    &  0.15 & KPNO \\
        2456086.841888 &   4.96    &  0.15 & KPNO \\
        2456088.949298 &  10.42    &  0.16 & KPNO \\
        2456091.888598 &  19.25    &  0.20 & KPNO \\ 
        2456138.759680 &  -3.40    &  0.16 &  Tull \\
        2456238.605381 &  31.34    &  0.18 &  Tull \\ \hline
	\end{tabular}
\end{table}

% . . . . . . . . . . . . . . . . . . . . . . . . . . . . . . . . . . . . . . 

\begin{table}[ht]
	\centering
	\caption{Kepler-1661 ELC fitted parameters}
	\label{tab:fit}
	
	\begin{tabular}{llll}
		\hline
        \hline
		Parameter & Best Fit & $1\sigma$ & unit \\ \hline 
		\textit{Binary Star} \\
        Time of Conjunction, $T_{c,b}$  &  -23.28180 & 0.00007 & BJD-2455000 \\
        Period, $P_b$                  &  28.162539 & 0.00005 & days \\
        $\sqrt{e_b}$ cos $\omega_b$    &  0.270     & 0.002 & ... \\
        $\sqrt{e_b}$ sin $\omega_b$    &  0.199     & 0.007 & ... \\
        Inclination, $i_b$             &  88.76     & 0.02 &  \degree \\
        \\
	Primary Mass, $M_1$            &  0.841    & 0.022 & $M_\odot$ \\
        Primary RV Semi-Amplitude $K$, $K_1$             &  17.30    & 0.04 & km s$^{-1}$ \\
        Primary Radius, $R_1$          &  0.762    & 0.010 & $R_\odot$\\
        Ratio of Radii, $R_1 / R_2$    &  2.756    & 0.05 & ... \\
        Primary Temperature, $T_1$     &  5100     & 100 & K \\
        Temperature Ratio, $T_2 / T_1$ &  0.71     & 0.03 & ... \\ 
        \\
        LD Primary \kepler $q_1$       &  0.93    & 0.15 & ... \\
        LD Primary \kepler $q_2$       &  0.05    & 0.16 & ... \\
        LD Secondary \kepler $q_{1}$   &  0.73    & 0.17 & ... \\
        LD Secondary \kepler $q_{2}$   &  0.93    & 0.24 & ... \\
        LD Primary MLO ($R$) $q_1$     &  0.52    & 0.20 & ... \\
        LD Primary MLO ($R$) $q_2$     &  0.33    & 0.22 & ... \\ \hline
        \textit{Planet} \\
        Mass, $M_p$                    &  17       & 12 & $M_\oplus$ \\
        Radius Ratio with Primary Star, $R_1/R_p$ &  21.50    & 0.33 & ... \\
        Time of Conjunction, $T_{c,p}$                   &  1007.1  & 0.4 & BJD-2455000 \\
        Period, $P_p$                  &  175.06   & 0.07 & days \\
        $\sqrt{e_p}$ cos $\omega_p$    &  0.092    & 0.023 & ... \\
        $\sqrt{e_p}$ sin $\omega_p$    &  0.219    & 0.009 & ... \\
        Inclination, $i_p$             &  89.464   & 0.012 & degree \\
        Nodal Longitude, $\Omega_p$    &  0.61     & 0.03 & degree \\ \hline
	\end{tabular}
% \tablecomments{Primary Temperature and Radius were given a prior of 5100 K $\pm$ 
% 100 K and 0.743 $\pm$ 0.042 for the ELC fit based on Spectral and Gaia analysis.}
\end{table}
% . . . . . . . . . . . . . . . . . . . . . . . . . . . . . . . . . . . . . . 

\begin{table}[ht]
	\centering
	\caption{Kepler-1661 System Parameters}
	\label{tab:deriv}
	\begin{tabular}{llll}
		\hline
        \hline
		Parameter & Best Fit & 1$\sigma$ & unit \\ \hline 
        \textit{Primary Star} \\
        Primary Mass, $M_1$              &  0.841    & 0.022 & $M_\odot$ \\
        Primary Radius, $R_1$            &  0.762    & 0.010 & $R_\odot$ \\ 
        Primary Temperature, $T_1$       &  5100     & 100   & K \\ \hline
        \textit{Secondary Star} \\
        Secondary Mass, $M_2$              &  0.262  & 0.005 & $M_\odot$ \\
        Secondary Radius, $R_2$            &  0.276  & 0.006 & $R_\odot$ \\ 
        Secondary Temperature, $T_2$       &  3585   & 167   & K \\ \hline
        \textit{Planet} \\
        Planet Mass, $M_p$                 &  17     &  12   & $M_\oplus$ \\
        Planet Radius, $R_p$               &  3.87   & 0.06  & $R_\oplus$ \\
        Average Density, $\rho_p$          &  1.6    & 1.1   & g cm$^{-3}$ \\ \hline
        \textit{Binary Orbit at BJD = }2,454,960 \\
        Period, $P_b$                      &  28.162539 &  0.00005 & days \\
        Time of Conjunction, $T_{c,b}$      &  -23.28180 & 0.00007 & BJD-2455000 \\
        Semimajor axis of Binary, $a_b$    &  0.187     & 0.002 & AU \\ 
        Eccentricity, $e_b$                &  0.112     & 0.002 & ... \\ 
        Argument of Periastron, $\omega_b$ &  36.4      & 1.1   & degree \\
        Inclination, $i_p$                 &  88.76     & 0.02 & degree \\ 
        Primary Impact Parameter, $b_1$    &  0.755     &  & ... \\ 
        Secondary Impact Parameter, $b_2$  &  0.862     &  & ... \\ \hline
        \textit{Planet Orbit at BJD = }2,454,960 \\
        Orbital Period, $P_p$                   &  175.06   & 0.06  & days \\
        Time of Barycenter Conjunction, $T_c,p$ &  1007.1   & 0.4 & BJD-2455000 \\
        Semimajor axis, $a_p$                   &  0.633    & 0.005 & AU \\ 
        Eccentricity, $e_p$                     &  0.057    & 0.005 & ... \\
        Argument of Periastron, $\omega_p$      &  67.1     & 5.0  & degree \\
        Inclination, $i_p$                      &  89.46    & 0.02 & degree \\
        Nodal Longitude, $\Omega_p$             &  0.61     & 0.03 & degree \\
        Mutual Inclination, $\Delta i$          &  0.93     & 0.02  &  degree \\ \hline
	\end{tabular}
\end{table}

% . . . . . . . . . . . . . . . . . . . . . . . . . . . . . . . . . . . . . . 

\begin{table}[ht]
	\centering
	\caption{Kepler-1661 Barycentric Cartesian Positions and Velocities at BJD = 2,454,960}
	\label{tab:cart}
	\begin{tabular}{llll}
		\hline
        \hline
Parameter             & Primary Star             &  Secondary Star          & Planet  \\ \hline
Mass, $M_\odot$       &  8.40853449093473149E-01 &  2.62347276214166703E-01 &  5.11075180194982723E-05 \\
$x$, \textit{AU}       &  3.53188589024455990E-02 & -1.13185968929754660E-01 & -7.73598863314166052E-02 \\ 
$y$, \textit{AU}       &  7.47539718007338670E-04 & -2.39687876159461882E-03 &  4.75985351132452317E-03 \\ 
$z$, \textit{AU}       &  3.46026030406624344E-02 & -1.11021524038963662E-01 &  5.96317715973122775E-01 \\ 
$V_x$, \textit{AU/day} & -6.32910801823011269E-03 &  2.02901512397735267E-02 & -2.37461133149127167E-02 \\
$V_y$, \textit{AU/day} &  1.34484593955837620E-04 & -4.30985153500305885E-04 & -2.74980447028744235E-04 \\
$V_z$, \textit{AU/day} &  6.22776497258325894E-03 & -1.99602373233647816E-02 & -2.42161959431956781E-03 \\ \hline
	\end{tabular}
\end{table}

% . . . . . . . . . . . . . . . . . . . . . . . . . . . . . . . . . . . . . . 
\begin{table}[ht]
	\centering
	\caption{Kepler-1661 Predicted Transits of the Planet Across 
the Primary Star}
	\label{tab:trans}
	\begin{tabular}{lllllll}
		\hline
        \hline
  & Date & & UT &  BJD - 2,455,000                     &  Impact Parameter                          &    Duration (days) \\ \hline
  2009 & May & ~3  & ~7:22:04.8 & $-45.193   ^{ +0.033  }_{-0.031    }$  &  $  -1.673  ^{+0.038  }_{-0.041   }$ & ... \\
  2009 & Oct & 20  & ~3:41:45.6 & $124.654   ^{ +0.023  }_{-0.023    }$  &  $  -1.486  ^{+0.027  }_{-0.040   }$ & ... \\
  2010 & Apr & ~8  & ~1:33:36.0 & $294.565   ^{ +0.015  }_{-0.017    }$  &  $  -1.319  ^{+0.029  }_{-0.024   }$ & ... \\
  2010 & Sep & 25  & ~1:11:08.2 & $464.5494  ^{ +0.0082 }_{-0.0104   }$  &  $  -1.149  ^{+0.020  }_{-0.016   }$ & ... \\
  2011 & Mar & 14  & ~2:57:15.8 & $634.6231  ^{ +0.0028 }_{-0.0058   }$  &  $  -0.9911 ^{+0.0099 }_{-0.0168  }$ & $  0.058  ^{+0.026 }_{ -0.019}$ \\
  2011 & Aug & 31  & ~7:13:09.1 & $804.8008  ^{ +0.0014 }_{-0.0019   }$  &  $  -0.8578 ^{+0.0086 }_{-0.0077  }$ & $  0.2186 ^{+0.0054}_{ -0.0065}$ \\
  2012 & Feb & 17  & 14:19:58.1 & $975.0972  ^{ +0.0013 }_{-0.0013   }$  &  $  -0.7489 ^{+0.0105 }_{-0.0067  }$ & $  0.3103 ^{+0.0033}_{ -0.0056}$ \\
  2012 & Aug & ~6  & ~0:10:56.6 & $1145.5076 ^{ +0.0019 }_{-0.0013   }$  &  $  -0.672  ^{+0.013  }_{-0.015   }$ & $  0.3724 ^{+0.0047}_{ -0.0072}$ \\
  2013 & Jan & 23  & 11:52:04.8 & $1315.9945 ^{ +0.0062 }_{-0.0042   }$  &  $  -0.649  ^{+0.025  }_{-0.020   }$ & $  0.3979 ^{+0.0069}_{ -0.0113}$ \\
  2013 & Jul & 12  & 23:30:46.1 & $1486.4797 ^{ +0.0092 }_{-0.0107   }$  &  $  -0.676  ^{+0.035  }_{-0.029   }$ & $  0.372  ^{+0.014 }_{ -0.013}$ \\
  2013 & Dec & 30  & ~8:45:36.0 & $1656.865  ^{ +0.017  }_{-0.012    }$  &  $  -0.755  ^{+0.043  }_{-0.035   }$ & $  0.308  ^{+0.019 }_{ -0.021}$ \\
  2014 & Jun & 18  & 14:32:38.4 & $1827.106  ^{ +0.018  }_{-0.026    }$  &  $  -0.871  ^{+0.050  }_{-0.041   }$ & $  0.210  ^{+0.032 }_{ -0.029}$ \\
  2014 & Dec & ~5  & 16:22:04.8 & $1997.182  ^{ +0.029  }_{-0.030    }$  &  $  -0.997  ^{+0.046  }_{-0.056   }$ & $  0.104  ^{+0.035 }_{ -0.048}$ \\
  2044 & Sep & ~4  & ~4:48:00.0 & $12862.7   ^{ +2.0    }_{-2.7      }$  &  $  -1.092  ^{+0.061  }_{-0.085   }$ & $  0.096  ^{+0.016 }_{ -0.054}$ \\
  2045 & Feb & 21  & 16:48:00.0 & $13033.2   ^{ +2.0    }_{-2.6      }$  &  $  -0.975  ^{+0.049  }_{-0.063   }$ & $  0.140  ^{+0.073 }_{ -0.059}$ \\
  2045 & Aug & 11  & ~4:48:00.0 & $13203.7   ^{ +1.9    }_{-2.4      }$  &  $  -0.900  ^{+0.064  }_{-0.049   }$ & $  0.229  ^{+0.049 }_{ -0.074}$ \\
  2046 & Jan & 28  & ~9:35:60.0 & $13373.9   ^{ +1.9    }_{-2.1      }$  &  $  -0.825  ^{+0.047  }_{-0.100   }$ & $  0.272  ^{+0.040 }_{ -0.104}$ \\
  2046 & Jul & 17  & 12:00:00.0 & $13544.0   ^{ +1.6    }_{-1.7      }$  &  $  -0.838  ^{+0.095  }_{-0.092   }$ & $  0.169  ^{+0.098 }_{ -0.075}$ \\
  2047 & Jan & ~3  & 14:24:00.0 & $13714.1   ^{ +1.3    }_{-1.7      }$  &  $  -0.910  ^{+0.127  }_{-0.077   }$ & $  0.128  ^{+0.091 }_{ -0.065}$ \\
  2047 & Jun & 22  & ~9:35:60.0 & $13883.9   ^{ +1.3    }_{-1.3      }$  &  $  -0.948  ^{+0.110  }_{-0.083   }$ & $  0.101  ^{+0.092 }_{ -0.052}$ \\
  2047 & Dec & ~9  & ~4:48:00.0 & $14053.7   ^{ +1.2    }_{-1.5      }$  &  $  -1.013  ^{+0.108  }_{-0.059   }$ & $  0.086  ^{+0.072 }_{ -0.052}$ \\
  2048 & May & 27  & ~0:00:00.0 & $14223.5   ^{ +1.0    }_{-1.4      }$  &  $  -1.050  ^{+0.078  }_{-0.080   }$ & $  0.061  ^{+0.068 }_{ -0.036}$ \\
  2048 & Nov & 12  & 14:24:00.0 & $14393.1   ^{ +1.0    }_{-1.3      }$  &  $  -1.133  ^{+0.087  }_{-0.060   }$ & $  0.089  ^{+0.047 }_{ -0.060}$ \\ \hline

	\end{tabular}
\tablecomments{The transits at 2011 Aug 31, 2012 Feb 17, and 2012 Aug 6 were observed during the \textit{Kepler} mission.}
\end{table}
% ------------------------------------------------------------------------------------

% ------------------------------------------------------------------------------------

\appendix
\section{Attempted Debiasing of the Primary Eclipses}\label{app:one}

In an attempt to correct the eclipse-depth bias, we measured the 
peak-to-peak amplitude of the starspot modulation in the normalized (but 
not detrended) light curve. The peak-to-peak amplitude in a sliding boxcar 
of width 50 days (roughly twice the stellar rotation period -- see Section \ref{sec:rot}) was used to provide slight smoothing of the variations. Figure 
\ref{fig:eclipse-depth} shows the light curve, the starspot modulation 
amplitude, and the actual measured depth of each primary eclipse. 
The starspot amplitude time series, $A(t)$, was then used to correct the 
usual normalized and detrended light curve $D(t)$ to produce a 
de-biased data:  
\begin{equation}
    D^\prime = D - A(D-1)
\end{equation} 
The amplitude correction term A ranges from 0.5\% to 2.5\%, and
the larger the starspot amplitude, the more the eclipse depth is 
decreased. 
The eclipse depths after this de-biasing no longer had a long-term tilt,
but small eclipse-to-eclipse variations were still present.

The result of using the de-biased light curve was a noticeable decrease 
in best-fit value for the planet mass. Yet the mass still remained 
abnormally high, $\sim$70 \Mearth, and the model still preferred 
a higher primary star mass than expected. 
So unfortunately this simple prescription for the de-biasing was 
insufficient to yield a satisfactory solution. The de-biasing
method was abandoned for the more straightforward method described in
Section \ref{sec:revised}, but not without exploring one more attempt to find
a method that gave a sensible planet mass.
For this approach we re-normalizd all the primary eclipses to same level
in flux, 0.9628 (a depth of 3.72\%) which is the average depth that was 
observed during the times of the least starspot activity.
By construction this removes sensitivity to the eclipse depth variations, 
and leaves only the eclipse timing variations as the source of constraining 
the planet's mass. This method has limitations, but it was useful as a 
check. The resulting parameter estimates agreed to within 1-$\sigma$ 
with the adopted method described in section \ref{sec:revised}. 


\begin{thebibliography}{}

\bibitem[Borkovits et al.(2011)]{Borkovits2011}
Borkovits, T., Csizmadia, S., Forg{\'a}cs-Dajka, E. \& 
Heged{\"u}s, T.
2011, \aap, 528, A53

\bibitem[Borkovits et al.(2015)]{Borkovits2015}
Borkovits, T., Rappaport, S., Hajdu, T. \& Sztakovics, J.
2015, \mnras, 448, 946-993

\bibitem[Bressan et al.(2012)]{Bressan2012} 
Bressan, A., Marigo, P., Girardi, L., et al.\ 
2012, \mnras, 427, 127 

\bibitem[Casagrande et al.(2010)]{2010A&A...512A..54C} Casagrande, L., Ram{\'\i}rez, I., Mel{\'e}ndez, J., et al.\ 2010, \aap, 512, A54

\bibitem[Chambers et al.(1996)]{Chambers1996} 
Chambers, J.~E., Wetherill, G.~W., \& Boss, A.~P.\ 1996, \icarus, 119, 261

\bibitem[Collins et al.(2017)]{Collins2017} 
Collins, K.A., Kielkopf, J.F., Stassun, K.G., 
\& Hessman, F.V.\ 2017, \aj, 153, 77

\bibitem[Dotter et al.(2008)]{Dotter2008}
Dotter, A., Chaboyer, B., Jevremovic, D., 
et al.\ 2008, \apjs, 178, 89

\bibitem[Doyle et al.(2011)]{Doyle_2011}
Doyle, L.~R., Carter, J.~A., Fabrycky, D.~C., 
et al.\ 2011, Science, 333, 1602

\bibitem[Eastman et al.(2010)]{2010PASP..122..935E} 
Eastman, J., Siverd, R., \& Gaudi, B.~S.\ 
2010, \pasp, 122, 935

\bibitem[Endl \& Cochran(2016)]{2016PASP..128i4502E} Endl, M. \& Cochran, W.~D.\ 2016, \pasp, 128, 094502

\bibitem[Gaia Collaboration et al.(2016)]{2016A&A...595A...1G} 
Gaia Collaboration,
Prusti, T., de Bruijne, J.H.J., et al.\ 2016, \aap, 595, A1

\bibitem[Gaia Collaboration et al.(2018)]{2018A&A...616A...1G} 
Gaia Collaboration,
Brown, A.~G.~A., Vallenari, A., et al.\ 2018, \aap, 616, A1

\bibitem[Gim{\'e}nez(2006)]{2006A&A...450.1231G} Gim{\'e}nez, A.\ 2006, \aap, 450, 1231

\bibitem[Hairer \& Hairer(2003)]{Hairer} Hairer, E. \& Hairer, M. \ 2003, Frontiers in Numerical Analysis (Durham, 2002) (Berlin: Springer) The code can be downloaded at http://www.unige.ch/math/folks/hairer

\bibitem[Hilditch(2001)]{2001icbs.book.....H} Hilditch, R.~W.\ 2001, An Introduction to Close Binary Stars

\bibitem[Holman \& Wiegert(1999)]{Holman1999}
Holman, M.J. \& Wiegert, P.A.
1999, \aj, 117, 621

\bibitem[Hut(1981)]{Hut81}
Hut, P. 1981, \aap, 99, 126

\bibitem[Kipping(2013)]{2013MNRAS.435.2152K} Kipping, D.~M.\ 2013, \mnras, 435, 2152

\bibitem[Kopparapu et~al.(2013a)]{Kopparapu_2013a}   
Kopparapu, R.~K.,  Ramirez, R.,  Kasting, J.~F.,
et al.\ 2013a, \apj, 765, 131

\bibitem[Kopparapu et~al.(2013b)]{Kopparapu_2013b}
Kopparapu, R.~K.,  Ramirez, R.,  Kasting, J.~F.,
et al.\ 2013b, \apj, 770, 82

\bibitem[Kostov et al.(2014)]{Kostov2014} 
Kostov, V.~B., McCullough, P.~R., 
Carter, J.~A., et al.\ 2014, \apj, 784, 14

\bibitem[Kostov et al.(2016)]{Kostov2016} Kostov, V.~B., Orosz, J.~A., Welsh, W.~F., et al.\ 2016, \apj, 827, 86

\bibitem[Kratter, \& Shannon(2014)]{Kratter2014} 
Kratter, K.~M., \& Shannon, A.\ 2014, \mnras, 437, 3727

\bibitem[Lam, \& Kipping(2018)]{Lam2018} 
Lam, C., \& Kipping, D.\ 2018, 
\mnras, 476, 5692

\bibitem[Li et al.(2016)]{Li2016} 
Li, G., Holman, M.~J., \& Tao, M.\ 
2016, \apj, 831, 96

\bibitem[Lissauer et al.(2011)]{Lissauer2011} 
Lissauer, J.~J., Fabrycky, D.~C., Ford, E.~B., et al.\ 
2011, \nat, 470, 53

\bibitem[Mandel, \& Agol(2002)]{2002ApJ...580L.171M} Mandel, K., \& Agol, E.\ 2002, \apjl, 580, L171

\bibitem[Mardling, \& Lin(2002)]{2002ApJ...573..829M} Mardling, R.~A., \& Lin, D.~N.~C.\ 2002, \apj, 573, 829

\bibitem[Mazeh et al.(2015)]{Mazeh2015}
Mazeh, T., Hoczer, T., Shporer, A. et al. 2015
2015, \apj, 800, 142 

\bibitem[McQuillan et al.(2013)]{2013MNRAS.432.1203M} McQuillan, A., Aigrain, S., \& Mazeh, T.\ 2013, \mnras, 432, 1203

\bibitem[Mudryk, \& Wu(2006)]{Mudryk2006} 
Mudryk, L.~R., \& Wu, Y.\ 2006, 
\apj, 639, 423

\bibitem[M\"uller \& Haghighipour(2014)]{Mueller_2014}
M\"uller, T.~W.~A. \& Haghighipour N. 2014, \apj, 782, 26

\bibitem[Orosz, \& Hauschildt(2000)]{2000A&A...364..265O} Orosz, J.~A., \& Hauschildt, P.~H.\ 2000, \aap, 364, 265

\bibitem[Orosz et al.(2012)]{Orosz2012} Orosz, J.~A., Welsh, W.~F., Carter, J.~A., et al.\ 2012, Science, 337, 1511

\bibitem[Orosz et al.(2019)]{Orosz2019} Orosz, J.~A., Welsh, W.~F., Haghighipour, N., et al.\ 2019, \aj, 157, 174

\bibitem[Pr{\v s}a et al.(2011)]{Prsa2011}
Pr{\v s}a, A.,  Batalha, N., Slawson, R.W., 
et al.\ 2011 \aj 141, 83

\bibitem[Quarles et al.(2018)]{Quarles2018} 
Quarles, B., Satyal, S., Kostov, 
V., et al.\ 2018, \apj, 856, 150

\bibitem[Ragozzine \& Wolf(2009)]{2009ApJ...698.1778R} Ragozzine, D., \& Wolf, A.~S.\ 2009, \apj, 698, 1778

\bibitem[Rein\& Liu(2012)]{Rein2012} 
Rein, H., \& Liu, S.-F.\ 2012, \aap, 
537, A128

\bibitem[Rein \& Spiegel(2015)]{Rein2015} 
Rein, H., \& Spiegel, D.~S.\ 
2015, \mnras, 446, 1424

\bibitem[Riello et al.(2018)]{Riello2018} 
Riello, M., De Angeli, F., Evans, 
D.~W., et al.\ 2018, \aap, 616, A3

\bibitem[Short et al.(2018)]{2018AJ....156..297S} Short, D.~R., Orosz, J.~A., Windmiller, G., et al.\ 2018, \aj, 156, 297

\bibitem[Skilling(2004)]{Skilling2004}
Skilling, J. 2004, in AIP Conf. Ser., Nested Sampling, 
ed. R. Fischer, R. Preuss, \& U. V. Toussaint 
(Melville, NY: AIP) 395

\bibitem[Slawson et al.(2011)]{Slawson2011}
Slawson, R.W., Pr{\v s}a A., Welsh, W.F., 
et al.\ 2011 \aj, 142,160S

\bibitem[Sutherland \& Kratter(2019)]{Sutherland2019} 
Sutherland, A.~P., \& Kratter, K.~M.\ 
2019, \mnras, 487, 3288

\bibitem[ter Braak \& Vrugt(2006)]{tbv2006} ter Braak, C. J. F., \& Vrugt, J. A. 2006, Statistics and Computing, 16, 239

\bibitem[Torres et al.(2006)]{Torres2010}
Torres, G., Andersen, J., \& Gim\'{e}nez, A.\ 
2010, \aapr, 18, 67

\bibitem[Tull et al.(1995)]{tull1995}
Tull, R.~G., MacQueen, P.~J., Sneden, C. \& Lambert, D.~L. 
1995, \pasp, 107, 251

\bibitem[Tull(1998)]{tull1998}
Tull, R.~G.
1998, Proc.\ SPIE,  3355, 387

\bibitem[Weiss \& Marcy(2014)]{Weiss2014}
Weiss, L.~M., \& Marcy, G.~W.\ 
2014, \apjl, 783, L6

\bibitem[Welsh et al.(2012)]{2012Natur.481..475W} Welsh, W.~F., Orosz, J.~A., Carter, J.~A., et al.\ 2012, \nat, 481, 475

\bibitem[Welsh et al.(2015)]{Welsh2015} Welsh, W.~F., Orosz, J.~A., Short, D.~R., et al.\ 2015, \apj, 809, 26

\bibitem[Welsh \& Orosz(2018)]{Welsh2018}
Welsh, W.F. and Orosz, J.A.
2018, in {\it Handbook of Exoplanets}, ISBN 978-3-319-55332-0,
Springer International Publishing AG, 34

\bibitem[Wittenmyer et al.(2005)]{Wittenmyer2005}
Wittenmyer, R.A., Welsh, W.F., Orosz, J.A., et al.
2005, \apj, 632, 1157


\end{thebibliography}
\end{document}